\documentclass[lettersize,journal]{IEEEtran}
\usepackage{amsmath,amssymb,amsfonts}
\usepackage{amsthm}
\usepackage{algorithmic}
\usepackage{algorithm}
\usepackage{array}
\newtheorem{lemma}{Lemma}
\newtheorem{definition}{Definition}
\usepackage[caption=false,font=scriptsize,labelfont=rm,textfont=rm]{subfig}
\usepackage{textcomp}
\usepackage{stfloats}
\usepackage{url}
\usepackage{verbatim}
\usepackage{booktabs}
\usepackage{graphicx}
\usepackage{cite}
\usepackage{multirow}
\usepackage{color}
\usepackage{xcolor}
\usepackage{setspace}

\begin{document}

\title{Fairness-aware Competitive Bidding Influence Maximization in Social Networks}

\author{Congcong Zhang, Jingya Zhou,~\IEEEmembership{Member,~IEEE,} Jin Wang, ~\IEEEmembership{Member,~IEEE,} Jianxi Fan, Yingdan Shi
        % <-this % stops a space
\thanks{Manuscript received 8 February 2023; revised 26 April 2023 and 31 May 2023; accepted 6 June 2023. This work was partially supported by the National Natural Science Foundation of China under Grants 61972272, 62072321, the Natural Science Foundation of the Jiangsu Higher Education Institutions of China under Grants 21KJA520008, 22KJA520007, Qinlan Project of Jiangsu Province, and Six Talent Peak Project of Jiangsu Province (XYDXX-084). \emph{(Corresponding author: Jingya Zhou.)}}
\thanks{Congcong Zhang, Jingya Zhou, Jin Wang, Jianxi Fan, and Yingdan Shi are with the School of Computer Science and Technology, Soochow University, Suzhou, Jiangsu 215006, China (email: cczhang\_suda@outlook.com; jy\_zhou@suda.edu.cn; wjin1985@suda.edu.cn; jxfan@suda.edu.cn; ydshi98@stu.suda.edu.cn)}
}

\markboth{IEEE TRANSACTIONS ON COMPUTATIONAL SOCIAL SYSTEMS,~Vol.~XX, No.~X, ~202X}%
{Shell \MakeLowercase{\textit{et al.}}: A Sample Article Using IEEEtran.cls for IEEE Journals}


\maketitle

\begin{abstract}
Competitive Influence Maximization (CIM) has been studied for years due to its wide application in many domains. Most current studies primarily focus on the micro-level optimization by designing policies for one competitor to defeat its opponents. Furthermore, current studies ignore the fact that many influential nodes have their own starting prices, which may lead to inefficient budget allocation. In this paper, we propose a novel Competitive Bidding Influence Maximization (CBIM) problem, where the competitors allocate budgets to bid for the seeds attributed to the platform during multiple bidding rounds. To solve the CBIM problem, we propose a Fairness-aware Multi-agent Competitive Bidding Influence Maximization (FMCBIM) framework. In this framework, we present a Multi-agent Bidding Particle Environment (MBE) to model the competitors' interactions, and design a starting price adjustment mechanism to model the dynamic bidding environment. Moreover, we put forward a novel Multi-agent Competitive Bidding Influence Maximization (MCBIM) algorithm to optimize competitors' bidding policies. Extensive experiments on five datasets show that our work has good efficiency and effectiveness.
\end{abstract}


% \vspace{-0.5em}
\section{Introduction}
\IEEEPARstart{A}s a fundamental task, Influence Maximization (IM) \cite{DBLP:journals/tkde/ChenZLSZZ22} has been investigated for nearly two decades owing to the continued prosperity of social networks, where social influence is used to quantify the overall impact of each individual user within the social network \cite{DBLP:journals/tweb/GongCHXHWF21}. IM has already been widely applied to many domains, e.g., viral marketing \cite{GhayooriN21, DBLP:journals/www/ZhouFWWL19}, social recommendation \cite{DhelimNAHM21, DBLP:journals/ijon/Wang00WX22, 10011155}, information diffusion \cite{ClarkP11, PhamDNDH17, ShiZSWZ20,DBLP:journals/ipm/xgsun23}, etc. The objective of IM is to identify a group of $K$ influential nodes (i.e., seed nodes) such that they can activate the maximum number of network nodes given a specific diffusion model. 

As a variant of the IM problem, Competitive Influence Maximization (CIM) considers a more complicated and relatively realistic scenario where multi-party with conflicting interests diffuse their influence simultaneously in the network \cite{carnes2007maximizing, hong2020efficient, zuo2022online}. However, the main preoccupation of existing studies is the micro-level optimization of a specific party's benefit \cite{bharathi2007competitive,li2015getreal,zuo2022online}. It may lead to a serious disparity across different parties and an unhealthy competitive atmosphere. Moreover, few studies consider the starting prices of influential nodes during budget allocation \cite{masucci2014strategic, varma2018marketing, ansari2019competitive}, while the starting price plays an important role in many realistic competitive scenarios. For example, many influential bloggers on Facebook and Twitter have a quotation for advertisements, and the competitors need to bid to strive for those bloggers. The bloggers will choose the highest bidder to promote its product. Similarly, in viral marketing, many competitors allocate budgets to bid for influential nodes that also have their own starting prices. Thus the flexible budget allocation policy is critical for the competitors to bid for influential nodes in the market.

The above two factors motivate us to simulate an even more realistic scenario by taking into account both multi-party benefits and budget allocation with starting prices.

\begin{figure}[t]
\label{figure:1}
    \centering
    \includegraphics[width=\linewidth]{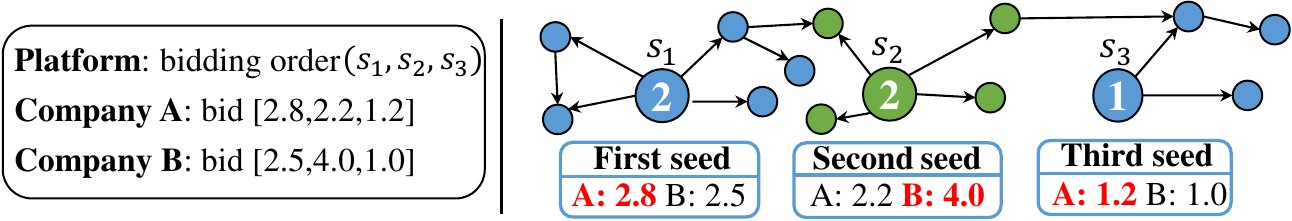}
    \caption{An example of the CBIM scenario in one bidding round. The number in each node indicates its starting price. Blue/green nodes are users activated by company $A$/$B$. Influential nodes $\{s_1, s_2, s_3\}$ are attributed to the platform.}
    \vspace{-0.3cm}
\end{figure}

\textbf{Example.} \emph{As shown in Fig. \ref{figure:1}, suppose there are two companies $A$ and $B$ with budgets $5$ and $5$ respectively, and three influential nodes $\{s_1, s_2, s_3\}$ with starting prices $[2,2,1]$ attributed to the platform, and the bidding is carried out in a fixed order $(s_1, s_2, s_3)$. Both companies intend to allocate their budgets to bid for $\{s_1, s_2, s_3\}$ so as to promote their new products, and only the winner needs to pay, which explains why the sum of each competitor's bidding prices exceeds its own budget. After one bidding round, company $A$ obtains $\{s_1,s_3\}$ as seed set and gets $10$ users activated, while company $B$ obtains $\{s_2\}$ as seed set and gets only $5$ users activated. There is a significant disparity in the promotion effect (e.g.,  $A$ actives twice as many users as $B$) though they spend the same cost (i.e., 4), which is unfair for company $B$.}

As illustrated by the Example, it is essential to consider fairness and budget allocation in the bidding scenario. However, if we only consider a single-round bidding scenario, it is difficult for the competitors to learn an effective bidding policy. Thus, we propose a novel Competitive Bidding Influence Maximization (CBIM) problem, where the competitors keep bidding the same seeds to optimize their bidding policies in multi-round bidding. In addition, the platform will adjust the starting prices of seeds in each bidding round to model the dynamic bidding environment. The goal of CBIM is to achieve a win-win situation where both the revenue of platform and the rewards of competitors are maximized while preserving fairness among competitors in the competitive bidding environment. 

\textbf{Challenges.} For the proposed CBIM problem, there are two main challenges:
(1) Simulate the interaction of the competitors. 
The competitors interact with both the bidding environment and themselves most critically. One competitor's change on its policy will inevitably affect the other competitors' policies, and vice versa. (2) Design an effective algorithm to optimize the competitors' policies to safeguard their benefits and preserve fairness among them. Since each competitor is self-interested, optimizing one competitor's interest may be harmful to other competitors.

Recently, there is a strong assumption in most statistical solutions of bidding optimization, which assumes the bidding policy is stationary \cite{masucci2014strategic, varma2018marketing, ansari2019competitive} (i.e., the competitors' policies do not change during bidding). To address it, we leverage Multi-agent Reinforcement Learning (MARL) to simulate the dynamic changes of the bidding environment and the competitors' interactions, and constantly optimize all competitors' bidding policies. Our motivation is twofold: (1) It enables each agent to strategically update its own policy to cope with the change of other agents. (2) It enables all bidding agents rather than an individual to maximize their own benefits.

% The purpose of this paper is to establish a framework to model the dynamic bidding process, and to achieve a win-win situation between the platform and competitors on the premise of fairness. 
The main contributions of our work are summarized below:
\begin{list}{\labelitemi}{\leftmargin=1em}
\setlength{\topmargin}{0pt}
\setlength{\itemsep}{0em}
\setlength{\parskip}{0pt}
\setlength{\parsep}{0pt}
\itemsep=0pt
    \item We propose a novel Competitive Bidding Influence Maximization (\textbf{CBIM}) problem and show that the problem is proven to be NP-hard.
    \item We propose a Fairness-aware Multi-agent Competitive Bidding Influence Maximization (\textbf{FMCBIM}) framework. The framework includes two modules: (1) A competition module where we develop a Multi-agent Bidding Particle Environment (\textbf{MBE}) for the competitors to bid and interact in a dynamic bidding manner. (2) A training module where we propose a practical Multi-agent Competitive Bidding Influence Maximization (\textbf{MCBIM}) algorithm for the competitors to optimize policies. The FMCBIM framework enables both the platform and the competitors to achieve a win-win situation on the premise of fairness.
    \item We conduct a great number of experiments based on five real-world datasets. The experimental results show that our proposed algorithm can effectively address the CBIM problem. {Moreover, the effectiveness of cooperation with the platform in our method shows that learning to cooperate in competition is more conducive to building a healthy competitive atmosphere in the market than pure competition.} 
\end{list}

The remainder of this paper is organized as follows. Section \ref{section:2} summarizes related work. Section \ref{section:3} formalizes the CBIM problem and prove its NP-hardness. Section \ref{section:4} elaborates our proposed FMCBIM framework. Section \ref{section:5} shows evaluation results. We conclude our work in Section \ref{section:6}.

% \vspace{-0.5em}
\section{Related Work}\label{section:2}
% \subsection{Competitive Influence Maximization} 
\textbf{CIM.} The CIM problem is firstly proposed by \cite{bharathi2007competitive} and \cite{carnes2007maximizing}. They address the CIM problem by assuming that the opponent's policy is known. 
Zuo et al. \cite{zuo2022online} introduce a variant of CIM, which is called Online Competitive Influence Maximization (OCIM). Its goal is to maximize the cumulative reward (i.e. influence spread) of one competitor with unknown influence probabilities on edges.
However, the scenarios in which the opponent's choice is known and fixed are unrealistic in the real world. Therefore, some researchers begin to explore the CIM problem by abandoning the above assumption.
% Lin et al. \cite{lin2015learning} treat the CIM problem without known competitors as a multi-round multi-party game in which the competitors learn their optimal policies via a learning-based framework. 
Li et al. \cite{li2015getreal} model the scenario of multi-party simultaneous selection of seed sets as a game, and put forward a framework that is more sensible compared with \cite{lin2015learning}, and it needs no information about the opponents.

The above studies are conducted from the perspective of competitors. Besides, some studies attempt to investigate from the host's (i.e., platform) perspective \cite{lu2013bang, zhu2018host, gao2020fair}. Lu et al. \cite{lu2013bang} aim to allocate, rather than identify, influential nodes to maximize all competitors' influence spreads from the platform's perspective while ensuring fairness among all competitors. Gao et al. \cite{gao2020fair} consider a scenario where the seed selection happens in event-based social networks (offline social events). They propose a RACE algorithm to select seeds for each event, which optimizes the solution by iteratively updating the seed selection probability via Cross Entropy (CE) method.

Different from the above studies, we take both perspectives of the platform and the competitors into account, and focus on achieving a win-win situation between the platform and the competitors on the premise of fairness. 
Moreover, few studies in the budget allocation task take into account the starting prices of influential nodes. The budget allocation task is first proposed by \cite{masucci2014strategic}. They use game theory to compute the Nash equilibrium of two competitors with equal budgets in the discrete voter model. 
Varma et al. \cite{varma2018marketing} consider the situation of two competitors with different budgets in the continuous voter model. 
Ansari et al. \cite{ansari2019competitive} consider a scenario where the competitors allocate budgets to compete for the given influential nodes, and they put forward a framework to compute the Nash equilibrium. 
{\color{black}Niknami et al. \cite{niknami2022competitive} propose a Max-Influence-Independent-Set (MIIST) algorithm to select the influential nodes, and then use the Colonel Blotto game to find the optimal investment policy (i.e., budget allocation scheme) for two competitors with equal budgets.} In comparison, our work can apply to a general multi-competitor scenario. Moreover, the consideration of the influential nodes' starting prices can avoid the low final transaction price of high-impact seed by adjusting the starting price over multi-round bidding. Consequently, existing approaches cannot well address our proposed CBIM problem in the competitive bidding environment. 

In CBIM, we not only need to treat competitors equally, but also balance the interests between competitors and the platform. Meanwhile, the competitors should cooperate with the platform to provide their observations for high-grade seeds. Therefore, considering multi-interest at the same time motivates us to build a framework, where we leverage MARL to explore the problem.

% \subsection{Multi-agent Reinforcement Learning}
\textbf{MARL.} In MARL, the type of task can be classified into fully cooperative setting, fully competitive setting and mixed setting by the designed reward structure \cite{gronauer2022multi}. 
In the fully cooperative setting \cite{2018counterfactual}, the agents having the identical reward function aim to maximize the mutual discounted reward by either communication protocol learning \cite{jiang2018learning} or parameters sharing \cite{foerster2016learning}. 
% Foerster et al. \cite{2018counterfactual} put forward a novel multi-agent actor-critic approach called COMA to learn decentralized policies for cooperative agents to address multi-agent credit assignment. 
In contrast, a competitive setting encourages agents to maximize their own  rewards and minimize the rewards of others at the same time \cite{lanctot2017unified, bansal2017emergent, lin2017multiagent}. 
% Bansal et al. \cite{bansal2017emergent} investigate the scenario in which agents learn motor skills by competing with simulated physics in a 3D world. 
The above MARL algorithms designed for fully cooperative or fully competitive scenarios are difficult to apply to our CBIM which aims to achieve a win-win situation. Unlike the settings of fully cooperative or fully competitive, the reward of an agent in the mixed setting is usually different and interrelated \cite{singh2018learning}. {\color{black}Lowe et al. \cite{lowe2017multi} adopt the framework of centralized training with decentralized execution (CTDE), which enables the incorporation of extra information to facilitate training, and propose MADDPG, a multi-agent deep deterministic policy gradient algorithm that can be adapted to different settings by redesigning the reward structure.} However, it is applied in a toy simulation environment and cannot handle dynamic bidding environments.

% \vspace{-0.5em}
\section{Problem Statement and Diffusion Model}\label{section:3}
\subsection{Problem Definition}\label{section:31}
In a social network denoted by a directed graph $G=(V, E)$, each node $v\in V$ represents a user and has its own activation threshold $\xi_v \in[0,1]$, each edge $e_{uv}$ (direction from $u$ to $v$) denotes that the node $u$ can activate node $v$ with an influence probability $w_{uv}\in(0,1]$. Let $C=\{c_1,...,c_k\}$ be a set of competitors with budgets $[b_1,...,b_k]$, and $\Theta=\{s_1,...,s_l\}$ be a set of seeds (influential users) attributed to the platform with initial starting prices $[p_1^1,...,p_l^1]$, where $k \geq 2$ and $l>0$. In the $t$-th bidding round, for each competitor $c_i$, $S_i^t$ denotes its seed set and $co_i^t$ denotes its cost. {\color{black}The main notations in this paper are shown in TABLE \ref{table:1}.}

\begin{table}[t]
\scriptsize
\setlength{\abovecaptionskip}{0cm}  
\setlength{\belowcaptionskip}{-0.2cm}
\renewcommand{\arraystretch}{0.9}
\caption{Main notations in this paper}
\begin{center}
\begin{tabular}{ll}
\toprule
\textbf{Notation}&\textbf{Description}\\
\midrule
$G = (V, E)$&A graph $G$ with node set $V$ and edge set $E$\\
$n$ & The number of nodes in $G$\\
$T_b$ & The number of total bidding rounds\\
$C$ & The set of competitors, $|C|=k$\\
$\Theta$& The set of bidding seeds, $|\Theta|=l$\\
$c_i,s_j$ & The $i$-th competitor and the $j$-th seed\\
$b_i$ & The budget of competitor $c_i$\\
$co_i^t$ & The cost of competitor $c_i$ in the $t$-th bidding round\\
$p_j^t$& The starting price of seed $s_j$ in the $t$-th bidding round\\
$S_i^t$& The seed set of competitor $c_i$ in the $t$-th bidding round\\
$GE$ & The fairness index of the competition\\
$CD_j^t$ & The contribution degree of seed $s_j$ in the $t$-th bidding round\\
$sc_j^t$ & The bidding result of seed $s_j$ in the $t$-th bidding round\\
$\sigma(S_i^t)$ & The reward of competitor $c_i$ in the $t$-th bidding round\\
$Inf(S_i^t)$ & The activated nodes set of seed set $S_i$ in the $t$-th bidding round\\
$\sum_{i=1}^k\sigma(S_i^t)$ &  The revenue of platform in the $t$-th bidding round\\
\bottomrule
\end{tabular}
\label{table:1}
\end{center}
\vspace{-0.5cm}
\end{table}

Before giving the definition of our problem, we first present some important concepts.
% \vspace*{-0.5\baselineskip}
\begin{definition}[Influence spread]
Given a seed set $S$ in a social network $G$ with $n$ nodes, the number of nodes activated by $S$ represents the influence spread and it can be calculated by $\sigma(S)=\sum_{j=1}^n \varsigma _j$, where $\varsigma _j=1$ when the node $v_j$ is activated by $S$, otherwise $\varsigma _j=0$.
\vspace*{-0.5\baselineskip}
\end{definition}
% \vspace*{-0.5\baselineskip}
\begin{definition}[Reward of competitor]
Given $k$ seed sets $\{S_1^t,...,S_k^t\}$ of $k$ competitors in the $t$-th bidding round, the reward of competitor $c_i$ is proportional to the benefit gained by $S_i^t$'s promotional activities, i.e., the influence spread of $S_i^t$. Therefore, the reward of competitor is defined as $\sigma(S_i^t)$.
\vspace*{-0.5\baselineskip}
\end{definition}
% \vspace*{-0.5\baselineskip}
\begin{definition}[Revenue of platform]
Given a set of bidding seeds $\Theta=\{s_1,..,s_l\}$ attributed to the platform, and $k$ seed sets $\{S_1^t,...,S_k^t\}$ of $k$ competitors, the revenue of platform in the $t$-th bidding round is defined as $\sum_{i=1}^k (\sigma(S_i^t))$.
\vspace*{-0.5\baselineskip}
\end{definition}
% \vspace*{-0.5\baselineskip}
\begin{definition}[Fairness index]
Given $k$ competitors with $[co_1^t,...,co_k^t]$, we choose Generalized Entropy \cite{mussard2003decomposition} as the fairness index to measure the fairness among competitors, i.e.,
% \vspace*{-0.5\baselineskip}
\begin{equation}\label{eq:1}
    GE = \frac{1}{k\cdot \omega(\omega-1)}\sum_{i=1}^k((\frac{R_i^t}{\bar{R^t}})^{\omega}-1),
% \vspace*{-0.5\baselineskip}
\end{equation}
where $GE$ denotes the fairness index, $R_i^t = \sigma(S_i^t)/co_i^t$ denotes the unit cost of competitor $c_i$, and $\bar{R^t} =\frac{1}{k}\sum_{i=1}^k R_i^t$ denotes the expected unit cost of all competitors in the $t$-th bidding round, $\omega$ is a constant. The closer $GE$ is to $0$, the fairer the competition is. (When $co_i^t=0$, we set $R_i^t=0$).
\vspace*{-0.5\baselineskip}
\end{definition}

To ensure smooth bidding, we follow the convention and present a series of business rules that all competitors have to obey and the competitors are not allowed to bid if they break the rules \cite{krishna2009auction}.
% \vspace*{-0.5\baselineskip}
\begin{definition}[Business rules]
 Suppose that there is a real-time bidding without reserve, given $l$ bidding items (seeds) with starting prices, $k$ bidders (competitors) with budgets, and a bidding master (platform), all competitors have to obey the following business rules: 
\begin{list}{\labelitemi}{\leftmargin=1em}
\setlength{\itemsep}{0em}
\setlength{\parskip}{0pt}
\setlength{\parsep}{0pt}
     \item No denying and no bargaining.
     \item Starting prices are confidential for the competitors.
     \item Seeds are auctioned off in a fixed order.
     \item Each competitor has only one chance to bid for all seeds.
     \item Rewards are returned only after all competitors have bid.
     \item The effective bidding price of the competitor must be higher than the starting price of the seed and should not exceed its remaining budget.
     \item The bidding payment mechanism employs the second price sealed-bid auction \cite{yuan2014empirical}.
\end{list}
\vspace*{-0.5\baselineskip}
\end{definition}

Based on all the above definitions, we formally define our problem as follows:

\textbf{CBIM Problem.} (Competitive Bidding Influence Maximization). In a social network $G$, there is a seed set $\Theta=\{s_1,..., s_l\}$ attributed to the platform with the initial starting prices $[p_1^1,...,p_l^1]$ and $k$ competitors with budgets. The total number of bidding rounds is $T_b$, and in each round, each competitor $c_i$ follows the business rules to bid $l$ seeds simultaneously, and then gets seed set $S_i^t$ at the end of current round. The starting prices of seeds are not fixed as the platform will adjust them based on the bidding results in the current round. The CBIM problem is to maximize the revenue of platform on the premise of fairness in the $T_b$ bidding rounds, i.e.,
% \vspace*{-0.5\baselineskip}
\begin{align}\label{eq:2}
    &\quad  \mathbb{S}^* = \underset{t\in [1,T_b],S_i^t\subseteq \Theta}{arg} max \sum_{i=1}^k \sigma(S_i^t), \nonumber \\
    s.t.\quad & (i) \, S_i^t\cap S_j^t =\varnothing, \, \forall i,\forall j \in [1,k], i\neq j, \nonumber \\
    &(ii) \, Inf(S_i^t)\cap Inf(S_j^t) =\varnothing, \, \forall i,\forall j \in [1,k], i\neq j, \nonumber \\
    &(iii) \, GE\leq \rho, \, \rho \in(0,1),
\vspace*{-0.5\baselineskip}
\end{align}
where $\mathbb{S}^*$ indicates the optimal solution, $Inf(S_i^t)$ denotes the node set activated by $S_i^t$, $\rho $ is the fairness threshold.
Constraint (i) indicates seed sets of any two competitors are disjoint. 
Constraint (ii) indicates the sets of activated nodes of any two competitors are non-overlapping.
Constraint (iii) indicates the fairness index should be less than a threshold to keep fairness.

As discussed, maximizing the revenue of platform is to maximize the total rewards of competitors. Meanwhile, the fairness among competitors should also be guaranteed. 
Therefore, maximizing the revenue of platform is equivalent to maximizing the reward of each competitor. Then we can achieve a win-win situation not only between the platform and competitors but also among the competitors. 

% \vspace{-0.5em}
\subsection{The Diffusion Model}
In our influence diffusion environment, the influence diffusion of each bidding round is independent and the influence spreads raised from different competitors' seed sets are assumed to propagate at one time.
In this paper, we illustrate our framework with the Competitive Linear Threshold (CLT) model \cite{lin2015learning} as an example. Because of the low coupling between the diffusion model and other modules, this framework is also applicable to the Competitive Independent Cascade (CIC) model \cite{wu2017scalable}. In addition, considering that we mainly focus on competitive bidding influence maximization instead of influence propagation process in this paper, we specifically concentrate on the situation that each node cannot be activated more than once. We leave the study of multi-time activation situations as future work.

% \vspace*{-0.5\baselineskip}
\begin{definition}[Competitive Linear Threshold (CLT) model]
It is a multi-competitor diffusion model. In each bidding round $t$, like LT model, node $v$ will be activated by seed set $S_i^t$ of competitor $c_i$ when $\sum_{u\in N^{in}_i(v)}w_{uv}>\xi_v$, where $w_{uv}$ is the probability that $v$ is activated by $u$, $N^{in}_i(v)$ is the set of $v$'s in-neighbors activated by $S_i^t$, and $\xi_v$ denotes $v$'s threshold of being activated. 
Node cannot be activated more than once.
When multiple competitors are eligible to activate  node $v$ simultaneously, it will be activated by $c_i$ who has the highest overall influence probability upon $v$, i.e., $\forall j\neq i, \sum_{u\in N_i^{in}(v)}w_{uv}>\sum_{u'\in N_j^{in}(v)}w_{u'v}$. Seeds with failed bids can no longer be activated by any nodes or activate other nodes. Finally, when no more nodes being activated or the total propagation time ends, the diffusion process terminates.
% \vspace*{-0.5\baselineskip}
\end{definition}

% \vspace{-0.5em}
\subsection{Properties of CBIM Problem}
\begin{lemma}\label{lemma:3.1}
The CBIM problem is NP-hard.
% \vspace*{-0.5\baselineskip}
\end{lemma}
\begin{proof}
To prove it, we reduce from Set Cover (SC) problem, a well-known NP-complete problem.
% i.e., CBIM $\succ$ SC.

\textbf{Set Cover (SC) Problem.} Given an integer $h>0$, a universal set $U=\{e_1,...,e_M\}$ and a collection of subsets $\boldsymbol{Y}=\{Y_1,...,Y_Z\}$, the SC problem asks if there are $h$ subsets whose union is $U$? 

Given a directed weighted graph $G=(V,E)$, $k$ competitors, $l$ bidding seeds, the total bidding rounds $T_b$, the seed set of each competitor in the $t$-th bidding round $S_i^t$, and an integer $\chi$, to simplify the analysis, here we assume $T_b=1$ and then we demonstrate that if SC  has a solution $\boldsymbol{Y}$ of size $h$, CBIM has a solution $\boldsymbol{S}$ of $k$ seed sets $S_1^1,...,S_k^1$ where $\sum_{i=1}^k \vert S_i^1\vert\leq l$, such that $\sum_{i=1}^{k}\sigma(S_i^1)\geq \chi $.

\textbf{Reduction.} Given two instances $P$ and $P'$ of SC problem defined as $U=\{e_1,...,e_M\}$, $\boldsymbol{Y}=\{Y_1,...,Y_Z\}$ and $h$, and $U'=\{e_1',...,e_{M'}'\}$, $\boldsymbol{Y}'=\{Y_1',...,Y_{Z'}'\}$ and $h'$, respectively, where $U=U'$, $\boldsymbol{Y}=\boldsymbol{Y}'$ and $h=h'$, then we make an instance $L$ for CBIM problem. Let $\vert V\vert=2(M+Z),k=2,\vert S_1^1\vert=h,\vert S_2^1\vert=h'$ and $\chi= 2(M+h)$. 
Elements $e_i\in U$($e_i'\in U'$) in $P$($P'$) are equivalent to nodes $v_i\in V$($v_i'\in V$) in $L$. The activation threshold is $0.5$ for both $v_i$ and $v_i'$. Subsets $Y_i\in \boldsymbol{Y}$($Y_i'\in \boldsymbol{Y}'$) in $P$($P'$) are equivalent to seeds $s_i\in V$($s_i'\in V$ ) in $L$. For each $e_i\in Y_j$($e_i'\in Y_j'$) in $P$($P'$), there is an edge connected from $s_j$($s_j'$) to $v_i$($v_i'$) with influence probability $1$ in $L$, i.e., $s_j$($s_j'$) can activate $v_i$($v_i'$) in $G$.
For the instance $L$, CBIM problem has a solution with double seed sets $S_1^1$ and $S_2^1$, including $h$ and $h'$ seeds, respectively, and $\sum_{i=1}^{2}\sigma(S_i^1)\geq 2(M+h)$, if and only if SC problem has a solution.

($\Leftarrow $): Suppose there are $h(h')$ subsets whose union equals $U(U')$. Since $h(h')$ subsets are equivalent to $h(h')$ seeds in $G$ and $M(M')$ elements $e_1,...,e_M(e_1',...,e_{M'}')$ in $U(U')$ are equivalent to $M(M')$ nodes $v_1,...,v_M(v_1',...,v_{M'}')$ activated by these $h(h')$ seeds, $\sum_{i=1}^k\sigma(S_i^1)=2(M+h)$ is no less than $\chi  = 2(M+h)$.

($\Rightarrow$): Suppose there are $h(h')$ seeds of competitor $c_1(c_2)$ and they activate $M(M')$ nodes $v_1,...,v_M(v_1',...,v_{M'}')$ such that $\sum_{i=1}^k\sigma(S_i^1)=\chi=2(M+h)$. Thus, there are $h(h')$ subsets whose union equals $U(U')$.

Therefore, the problem of CBIM is NP-hard.
\end{proof}

\begin{lemma}\label{lemma:3.2}
The CBIM problem is monotone.
% \vspace*{-0.5\baselineskip}
\end{lemma}
\begin{proof} Given a graph $G=(V,E)$, supposing there are two competitors $c_1$ and $c_2$ bidding for a seed set $\Theta$ in $T_b$ bidding rounds, then the objective function of CBIM can be converted as $\sigma(S_1^t)+\sigma(S_2^t)$,
where $S_1^t\subseteq \Theta$ is the seed set of competitor $c_1$ in the $t$-th bidding round, $S_2^t\subseteq \Theta$ is the seed set of competitor $c_2$. As \cite{bharathi2007competitive} shows in the proof of lemma 1, $\sigma(S_1^t)$ and $\sigma(S_2^t)$ are both monotone. According to the addition principle, we have $\sigma(S_1^t)+\sigma(S_2^t)\leq \sigma(T_1^t)+\sigma(T_2^t)$ for subsets $S_1^t\subseteq T_1^t\subseteq V$, $S_2^t\subseteq T_2^t\subseteq V$. Since the influence diffusion in each bidding round is independent, the above conclusions are valid in all bidding rounds. Thus the objective function of CBIM is monotone.
\end{proof}

% \vspace{-0.5cm}
\subsection{Illustrative Example}

\begin{figure*}[t]
\centering
\subfloat[An example graph]
{
\begin{minipage}[b]{.3\linewidth}
        \centering
        \includegraphics[width=.9\linewidth]{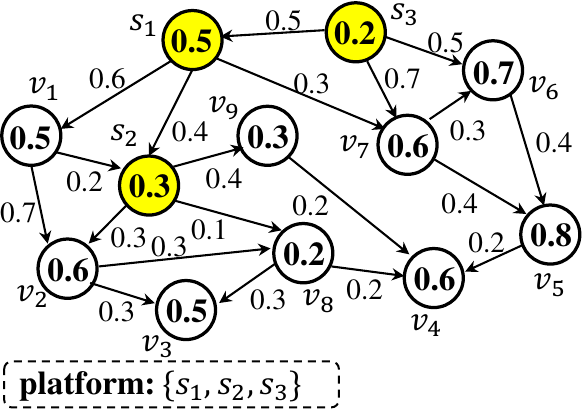}
    \end{minipage}
\label{figure:2a}
}
\subfloat[A possible result of the first round]
{
 \begin{minipage}[b]{.3\linewidth}
\centering
\includegraphics[width=.9\linewidth]{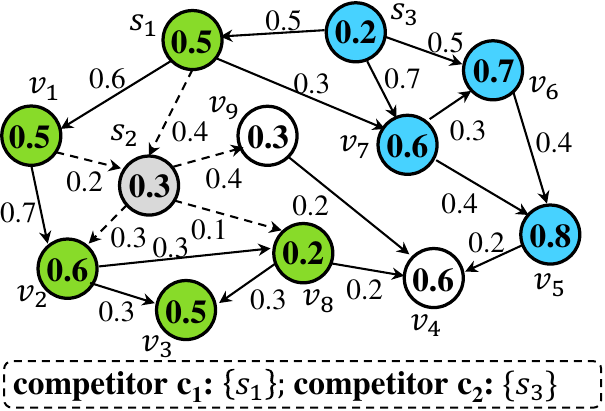}
\end{minipage}
\label{figure:2b}
}
\subfloat[A possible result of the second round]
{
 	\begin{minipage}[b]{.3\linewidth}
        \centering
        \includegraphics[width=.9\linewidth]{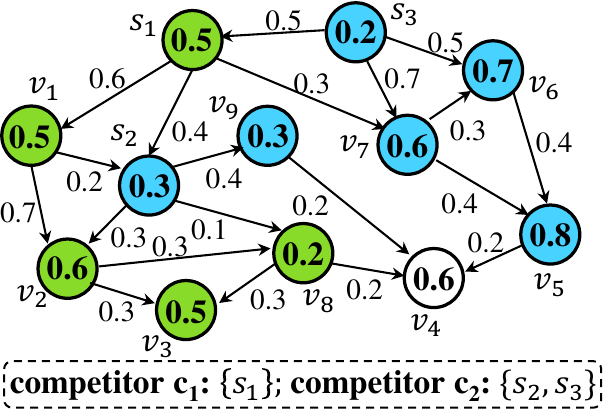}
    \end{minipage}
    \label{figure:2c}
}
\caption{An illustrative example of CBIM. Yellow nodes $\{s_1,s_2,s_3\}$ are bidding seeds. Green nodes are users activated by $c_1$. Blue nodes are users activated by $c_2$. Grey node is the seed of bidding failure. Dashed arrows indicate blocked edges.}
\label{figure:2}
\vspace{-0.3cm}
\end{figure*}

As Fig. \ref{figure:2} shows, in this illustrative social graph, 
% where the number in each node is the activation threshold, and the number on the directed edge is the influence probability.
 the numbers on nodes and directed edges denote the activation thresholds and influence probabilities, respectively.
We consider a scenario where two competitors $c_1$ and $c_2$ with budgets $[b_1=3, b_2=3]$ plan to find influential nodes to promote their new products, in order to achieve the maximum number of activated users who buy their products. $\Theta=\{s_1, s_2, s_3\}$ is the seed set with starting prices $[p_1^1=1, p_2^1=1, p_3^1=1]$ attributed to the platform and the seeds' bidding order is $(s_1, s_2, s_3)$. We set the fairness threshold as $\rho =0.1$ and the constant as $\omega=2$.

Fig. \ref{figure:2a} shows the initial graph. A possible result of the first bidding round is shown in Fig. \ref{figure:2b}: $c_1$ bids for $\Theta$ with bidding prices $[2.0, 0.0, 1.0]$ and gets $\{s_1\}$ as seed set and achieves $5$ activated users (including seed $s_1$ itself). $c_2$ bids for $\Theta$ with bidding prices $[1.5, 0.0, 2.0]$ and gets $\{s_3\}$ as seed set and achieves $4$ activated users. $s_2$ is not auctioned off successfully.
The rewards of both $c_1$, $c_2$ and the revenue of platform are as follows: $\sigma(S_1^1)=5$ and $\sigma(S_2^1)=4$ and $\sigma(S_1^1)+\sigma(S_2^1)=9$. Although the bidding results satisfy the fairness constraint, i.e., $GE=0.059<0.1$, the revenue of platform is not maximized.

Then the platform adjusts the seeds' starting prices to $[p_1^2=1.1, p_2^2=0.9, p_3^2=1.1]$ based on the bidding results illustrated in Fig. \ref{figure:2b}. A possible result of the second bidding round is shown in Fig. \ref{figure:2c}: $c_1$ bids for seeds with bidding prices $[2.0, 1.0,0.0]$ and gets $\{s_1\}$ as seed set and achieves $5$ activated users. $c_2$ bids for seeds with bidding prices $[1.8,1.5,1.5]$ and gets $\{s_2,s_3\}$ as seed set and achieves $6$ activated users. The results are as follows: $\sigma(S_1^2)=5$ and $\sigma(S_2^2)=6$ and $\sigma(S_1^2)+\sigma(S_2^2)=11$. 
The bidding results satisfy the fairness constraint, i.e., $GE=0.059<0.1$ and then the revenue of platform is further optimized.

% \vspace{-0.5em}
\section{Methodology}\label{section:4}
\begin{figure*}[htbp]
    \centering
    \includegraphics[width=0.8\linewidth]{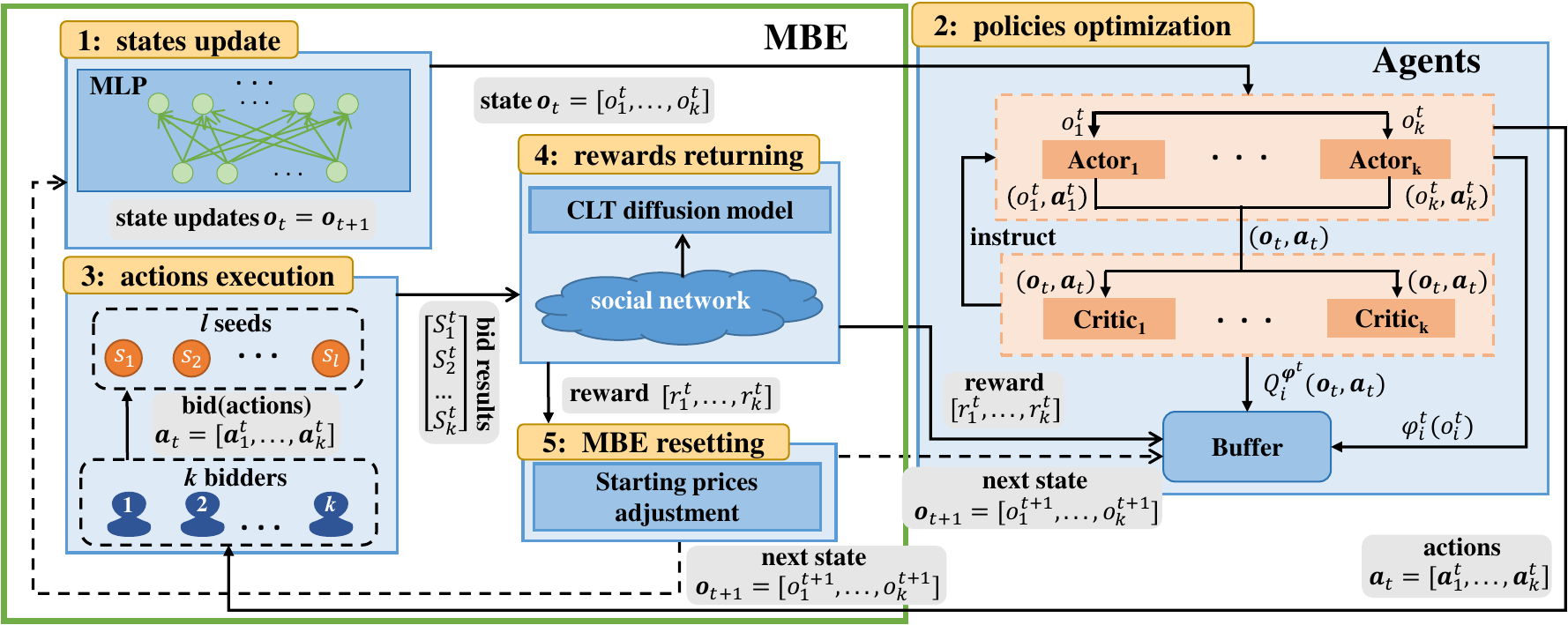}
    \caption{The overview of FMCBIM framework. Solid lines indicate data flows in the current bidding round while dotted lines indicate data flows input to the next bidding round.}
    \label{figure:3}
\vspace{-0.3cm}
\end{figure*}

We put forward a Fairness-aware Multi-agent Competitive Bidding Influence Maximization (FMCBIM) framework to deal with the CBIM. As shown in Fig. \ref{figure:3}, FMCBIM makes up of two modules: (1) A competition module, in which we present a bidding environment MBE for agents (competitors) to compete and interact. (2) A training module, in which we leverage the MARL technique for agents to optimize their own policies. 
During the training period, the agents keep interacting with each other and MBE to learn the connection among actions, states, and rewards, and then update policies for choosing actions independently.
In addition, the platform resets the MBE for the next round by adjusting the seeds' starting prices at the end of current round. 
To develop the FMCBIM framework, we first introduce MBE, the bidding environment, and then present the details of policy optimization.

% \vspace{-0.4cm}
\subsection{Multi-agent Bidding Particle Environment}
In CBIM, since the competitors compete with each other but cooperate with the platform, we develop a mixed virtual environment MBE based on MPE \cite{lowe2017multi} to model the competitors' interactions in bidding. In MBE, there are 3 kinds of objects, i.e., the platform, the seeds, and the competitors. We suppose that the competitors' budgets are predefined. The platform first identifies the influential nodes as bidding seeds in the network and sets the initial starting prices of seeds before training as:
% \vspace*{-0.5\baselineskip}
\begin{equation}
    p_j^1 = (\sum_{i=1}^k b_i)/l,\quad \forall j\in[1, l],
% \vspace*{-0.5\baselineskip}
\end{equation}
where $p_j^1$ is the initial starting price of seed $s_j$, $\sum_{i=1}^k b_i$ is the total budgets of all competitors. Since identifying influential nodes is not the focus of our work, {\color{black}we choose the degree heuristic algorithm \cite{jung2012irie} to select seeds}. Then the competitors bid over $l$ seeds to strive for their seed sets. After one bidding round, the platform returns rewards (i.e., the influence spreads under the CLT diffusion model) to the competitors, and adjusts the seeds' starting prices according to the bidding results of the current bidding round.

As we know, the bidding scenario in CBIM is a repeated process where the platform adjusts the seeds' starting prices by their influence spreads in each bidding round. The bidding results of seeds in each round are thus not independent. For the platform, the adjustment mechanism of seeds' starting prices affects the competitiveness of seeds during bidding, which in turn affects the association established between the platform and competitors. For each competitor, myopically utilizing a fixed bidding policy may not be able to cope with the dynamic and changeable bidding environment. On the contrary, 
% the competitors should consider the impact of the seed bidding results for the current round in the long term. 
in the long term, the competitors should take into account the potential impact of the seed bidding results at the current round. Therefore, it is significant for us to design a reasonable adjustment mechanism over starting prices.

We define $CD_j^t=\frac{\sigma(s_j^t)}{\sigma(\Theta)^t/l}$ as the seed $s_j$'s contribution degree to the total influence spreads of $l$ seeds in the $t$-th bidding round, where $\sigma(s_j^t)$ denotes the influence spread of seed $s_j^t$, and $\sigma(\Theta)^t/l$ denotes the average influence spread of $l$ seeds. Then considering the competitors' demands of gaining high influence spreads, we design the following adjustment principles for starting prices: (1) Depress the starting prices of bidding failed seeds. (2) Remain the starting prices unchanged for those seeds with successful bidding but lower contribution degree. (3) Raise the starting prices for those seeds with successful bidding and higher contribution degree. The adjustment formula is given as follows:
% \vspace*{-0.5\baselineskip}
\begin{equation}
    \label{eq:4}
    p_j^{t+1}=\left\{\begin{array}{l}
    p_j^t\cdot (1-\kappa),\quad if \, sc_j^t=0, \\
    p_j^t,\quad if\, sc_j^t=1 \wedge  CD_j^t <1,  \\
    p_j^t\cdot min(1+\kappa, CD_j^t), \quad if \, sc_j^t=1 \wedge CD_j^t \geq 1, 
    \end{array}\right.  
\end{equation}
where $p_j^{t+1}$ is the starting price of seed $s_j$ in the $(t+1)$-th bidding round, the value of $sc_j^t$ is $1$ if seed $s_j$ is bid successfully in the $t$-th bidding round, otherwise it is $0$, and $\kappa\in (0,1)$ is a parameter.

% \vspace{-0.4cm}
\subsection{Policy Optimization with MARL}
As a machine learning paradigm, MARL is primarily based on Markov Game (MG). In MARL, multi-agent simultaneously interact with each other in a shared environment. Their goal is to learn the optimal policy and get the maximum expected rewards from a long-term perspective. MARL defines the iterative process as a MG and utilizes dynamic programming to address the optimization problem for each agent. 

\subsubsection{Markov Game}
We formulate the bidding scenario in CBIM as an MG, where $k$ agents on behalf of $k$ competitors bid for $l$ seeds attributed to the platform. A MG is denoted by a group of all agents' possible states $\boldsymbol{O}$, and a group of actions $\boldsymbol{A}_1,...,\boldsymbol{A}_k$ where $\boldsymbol{A}_i$ indicates the agent $i$'s action space. 
The actions of each agent $i$ are calculated by using a policy $\pi_i:\boldsymbol{O}_i\to\boldsymbol{A}_i$ according to the $t$-th bidding round state $o_i^t$, and the action here is a bidding price $a_{ij}\in\boldsymbol{A}_i$, indicating that agent $i$ bids for seed $s_j$ with price $a_{ij}$.
After the action execution, agent $i$ transfers to the next state $o_i^{t+1}$ according to the state transition function $\boldsymbol{\Gamma} :\boldsymbol{O}\times\boldsymbol{A}_1\times...\times\boldsymbol{A}_k\mapsto\Omega(\boldsymbol{O})$, where $\Omega(\boldsymbol{O})$ denotes the set of probability distributions on the state space. Each agent $i$ owns an associated reward function $r_i:\boldsymbol{O}\times\boldsymbol{A}_1\times...\times\boldsymbol{A}_k\mapsto\boldsymbol{R}$ that provides a reward $r_i^t$, i.e., influence spread in the $t$-th bidding round.
We present detailed descriptions of several important concepts.
% in our game settings including agent, state, action, reward, and transition for clarity.

\textbf{Agent}: In MBE, $k$ competitors are denoted as $c_1,...,c_k$ and $l$ seeds are denoted as $s_1,...,s_l$. Each bidding round $t$ is launched by the platform and all agents bid over these seeds based on their starting prices $[p_1^t, ...,p_l^t]$. During bidding, all agents have to obey the business rules defined in Sec. \ref{section:31}.

\textbf{State}: The agent $i$'s remaining budget $rb_i^t$ in the $t$-th bidding round can be used to characterize its budget spent status for the next round. Meanwhile, the variation tendency on agents' effective bidding prices can indirectly reflect the changing trend of seeds' starting prices with the contribution degree. Therefore, we define the state $o_i^t$ of each agent $i$ in the $t$-th bidding round as $o_i^t=[\boldsymbol{g}_i^{t-1}, rb_i^{t-1}]$, where $\boldsymbol{g}_i^{t-1}=[g_{i1}^{t-1},...,g_{il}^{t-1}]$ indicates the effective bidding prices of agent $i$ in the $(t-1)$-th bidding round, $g_{ij}^{t-1}$ indicates the effective bidding price of agent $i$ towards seed $s_j$ and $g_{ij}^{t-1}=0$ when the bidding price is noneffective. 
{\color{black}Furthermore, to ensure fairness among competitors, we adopt the CTDE paradigm, in which the platform collects information from all competitors and uses it to update a joint policy. Subsequently, the platform instructs the decentralized competitors to learn their bidding policies.} In this paper, we provide all agents' states $\boldsymbol{o}_t=[o_1^t,...,o_k^t]$ as the extra information for agents to train their critic function.

\textbf{Action}: 
It is difficult for agents to explore the action space while the action is designed as a certain value. Therefore, we map the probabilities calculated by the neural network over a probability distribution as actions. Furthermore, we add the Gaussian noise to actions in order to help agents explore the environment more effectively. Unlike stochastic policy gradient (SPG) that outputs an action probability distribution and selects the action with the maximum probability, we output a specific action probability directly. In this paper, considering that many human behaviors are regular, we assume that the human bidding behavior satisfies normal distribution here according to Center Limit Theorem. For each bidding seed $s_j$, agent $i$ has an action $a_{ij}\in [0,b_i]$ (i.e., bidding price) and we denote the agent $i$'s actions in the $t$-th bidding round as $\boldsymbol{a}_i^t=[a_{i1}^t,...,a_{il}^t]$. Note that the actions of agent $i$ could be $\boldsymbol{a}_i^t=[0,...,0]$ in theory, indicating that the agent $i$ would like to skip this bidding round. However, there is no incentive for agents to skip one or more rounds, as they aim to maximize their personal rewards by learning bidding policies from past bidding experiences.  

\textbf{Reward and Transition}: Each agent $i$ aims to maximize its own total expected reward $r_i=\sum_{t=0}^{T_b}\gamma ^{t} r_i^t$ in the long run, where $\gamma \in(0,1]$ refers to a discount factor. The platform returns the reward $r_i^t$ to each agent $i$ only after all seeds have been auctioned out. After the $t$-th bidding round, agent $i$ updates its state from $o_i^t$ to $o_i^{t+1}=[\boldsymbol{g}_i^t, rb_i^t]$ with the changing of features $\boldsymbol{g}_i^t$ and $rb_i^t$.
% the state information of agent $i$ will update with the change of features $\boldsymbol{g}_i^t$ and $rb_i^t$, i.e.,  agent $i$ updates its state from $o_i^t$ to $o_i^{t+1}=[\boldsymbol{g}_i^t, rb_i^t]$.

\subsubsection{MARL in CBIM}
Considering that actions are taken from a continuous space, so we utilize the deterministic policy gradient (DPG) to optimize the bidding policy. In MARL, agent $i$'s $Q$ function is called critic, i.e.,
% \vspace*{-0.5\baselineskip}
\begin{equation}
    Q_i^{\boldsymbol{\pi}_t}(\boldsymbol{o}_t,\boldsymbol{a}_t)=\mathbb{E}_{\boldsymbol{\pi}_t,\boldsymbol{\Gamma}}[\sum_{j=1}^{t}\gamma^j r_i^j\vert \boldsymbol{o}_1=\boldsymbol{o}_t,\boldsymbol{a}_t],
    \label{eq:5}
% \vspace*{-0.5\baselineskip}  
\end{equation}
where $\boldsymbol{\pi}_t=\{\pi_1^t,...,\pi_k^t\}$ and $\boldsymbol{a}_t=[\boldsymbol{a}_1^t,\boldsymbol{a}_2^t,...,\boldsymbol{a}_k^t]$ are joint policy and joint action of all agents. $\boldsymbol{o}_1=[o_1^1,...,o_k^1]$ is the initial state vector. MARL exploits temporal difference recursive relationship with the state $\boldsymbol{o}_{t+1}=[o_1^{t+1},...,o_k^{t+1}]$ and action $\boldsymbol{a}_{t+1}$ to solve the optimal action decision in the $(t+1)$-th bidding round known as the Bellman equation:
% \vspace*{-0.3\baselineskip}
\begin{equation}
\begin{split}
    Q_i^{\boldsymbol{\pi}_t}(\boldsymbol{o}_t,\boldsymbol{a}_t)
    &=\mathbb{E}_{r,\boldsymbol{o}_{t+1}}[r(\boldsymbol{o}_t,\boldsymbol{a}_t)\\&+\gamma\mathbb{E}_{\boldsymbol{a}_{t+1}\sim\boldsymbol{\pi}_t}[Q_i^{\boldsymbol{\pi}_t}(\boldsymbol{o}_{t+1},\boldsymbol{a}_{t+1})]].
    \label{eq:6}
\end{split}
% \vspace*{-0.5\baselineskip}
\end{equation}

The policy for agent $i$ is a deterministic mapping function $\varphi _i(\cdot)$ from state $o_i^t$ to actions $\boldsymbol{a}_i^t$ with a parameter $\theta_i^{\varphi^t}$, and $\varphi _i(\cdot)$ is usually called actor, i.e.,
\begin{equation}
    \boldsymbol{a}_i^t = \varphi _i^t(o_i^t) = \varphi _i^t([\boldsymbol{g}_i^{t-1},rb_i^{t-1}]).
    \label{eq:7}
\end{equation}

We substitute Eq. (\ref{eq:7}) into Eq. (\ref{eq:6}) and obtain:
% \vspace*{-0.3\baselineskip}
\begin{equation}
    \begin{split}
        Q_i^{\boldsymbol{\varphi }_t}(\boldsymbol{o}_t&,\boldsymbol{a}_1^t,...,\boldsymbol{a}_k^t)
        =\mathbb{E}_{r,\boldsymbol{o}_{t+1}}[r(\boldsymbol{o}_t,\boldsymbol{a}_1^t...,\boldsymbol{a}_k^t)\\
        &+\gamma Q_i^{\boldsymbol{\varphi}_{t+1}}(\boldsymbol{o}_{t+1},\varphi _1^{t+1}(o_1^{t+1}),...,\varphi _k^{t+1}(o_k^{t+1}))],
        \label{eq:8}
    \end{split}
% \vspace*{-0.5\baselineskip}
\end{equation}
where $\boldsymbol{\varphi_t }=\{\varphi _1^t,...,\varphi _k^t\}$ is the joint deterministic policy of all agents.

We gradually update agent $i$'s $Q_i^{\boldsymbol{\varphi}_t }$ and $\varphi _i^t(o_i^t)$ independently. Concretely, 
% the critic $Q_i^{\boldsymbol{\varphi}_t }$ with parameter $\theta_i^{Q^t}$ is trained by minimizing loss $L(\theta_i^{Q^t})$ defined as follows:
we train the critic $Q_i^{\boldsymbol{\varphi}_t }$ by means of loss $L(\theta_i^{Q^t})$ minimization:
% \vspace*{-0.3\baselineskip}
\begin{equation}
    L(\theta_i^{Q^t})=\mathbb{E}_{\boldsymbol{o}_t,\boldsymbol{a}_t,r_i^t,\boldsymbol{o}_{t+1}}[(y_t-Q_i^{\boldsymbol{\varphi }_t}(\boldsymbol{o}_t,\boldsymbol{a}_1^t,...,\boldsymbol{a}_k^t))]^2,
    \label{eq:9}
\end{equation}
\begin{equation}
    y_t=r_i^t+\gamma Q_i^{\boldsymbol{\varphi }_{t+1}}(\boldsymbol{o}_{t+1},\varphi _1^{t+1}(o_1^{t+1}),...,\varphi _k^{t+1}(o_k^{t+1})),
    \label{eq:10}
\end{equation}
where $\boldsymbol{\varphi }_{t+1}=\{\varphi _1^{t+1},...,\varphi _k^{t+1}\}$ is the target policy with delayed parameter $\theta_i^{\boldsymbol{\varphi} _{t+1}}$, $Q_i^{\boldsymbol{\varphi}_{t+1}}$ is the target critic with delayed parameter $\theta_i^{Q^{t+1}}$, and $(\boldsymbol{o}_t, \boldsymbol{a}_t,r_i^t,\boldsymbol{o}_{t+1})$ is a transition tuple stored in the replay memory $D$, where $\boldsymbol{a}_t=[\boldsymbol{a}_1^t,...,\boldsymbol{a}_k^t]$. The policy of each agent $\varphi _i^t$ with parameter $\theta_i^{\boldsymbol{\varphi}^t }$ is trained by
% \begin{equation}
% \vspace*{-0.5\baselineskip}
\begin{align}\label{eq:11}
    \nabla_{\theta_i^{\boldsymbol{\varphi}^t }}&\textit{J}(\varphi _i^t)= \\
    &\mathbb{E}_{\boldsymbol{o}_t}[\nabla_{\theta_i^{\boldsymbol{\varphi}^t }}\varphi^t _i(o_i^t)\nabla_{\boldsymbol{a}_i^t}Q_i^{\boldsymbol{\varphi}_t}(\boldsymbol{o}_t, \boldsymbol{a}_1^t,...,\boldsymbol{a}_k^t)|_{\boldsymbol{a}_i^t=\varphi _i^t(o_i^t)}]. \nonumber
% \vspace*{-0.5\baselineskip}
\end{align}
% \end{equation}

\begin{algorithm}[tb]
\footnotesize
\renewcommand{\algorithmicrequire}{\textbf{Input:}}%¸ü¸ÄÊäÈëÃû³Æ
\renewcommand{\algorithmicensure}{\textbf{Output:}}%¸ü¸ÄÊä³öÃû³Æ
\caption{MCBIM}
\begin{algorithmic}[1]\label{algorithm:1}
    \REQUIRE Social network $G$, the number of competitors $k$, the budgets of competitors $[b_1,...,b_k]$, the set of bidding seeds $\Theta=\{s_1,...,s_l\}$, the number of iterations $N$, the number of bidding rounds $T$ in one iteration, initial states of all agents $\boldsymbol{o_1}=[o_1^1,...,o_k^1]$, target function update parameter $\tau$. 
    \ENSURE The maximum revenue of platform $\sum_{i=1}^k \sigma(S_i)$.
    \STATE For $\forall$ agent $i$, initialize critic $Q_i^{\boldsymbol{\varphi}_1}(\boldsymbol{o}_1,\boldsymbol{a}_1^1,...,\boldsymbol{a}_k^1|\theta_i^{Q^1})$, actor $\varphi _i^1(o_i^1|\theta_i^{\varphi^1})$, target critic $Q_i^{\boldsymbol{\varphi}_2}$ with $\theta_i^{Q^2} \leftarrow \theta_i^{Q^1}$, target policy $\varphi _i^2$ with $\theta_i^{\boldsymbol{\varphi} ^2}\leftarrow \theta_i^{\boldsymbol{\varphi}^1 }$;
    \STATE Initialize the replay buffer $D$;
    \FOR{$iteration = 1 \to N$}
        \item Initialize a defined bidding environment $\Re$ for action exploration;
        \item For $\forall$ agent $i$, receive an initial state $o_i^1$;
        \FOR{$t = 1\to T$}
            \STATE For $\forall$ agent $i$, compute $\boldsymbol{a}_i^t$ by $o_i^t$ with Eq. (\ref{eq:7}) and add them into $\Re$\, and get seed set $S_i^t$;
            \STATE For $\forall$ agent $i$, save reward $r_i^t$ of $S_i^t$, update state to new state $o_i^{t+1}$;
            \STATE For $\forall$ seed $s_j \in \Theta$, compute $p_j^{t+1}$ with Eq. (\ref{eq:4});
            \STATE Merge states as $\boldsymbol{o}_t=[o_1^t,...,o_k^t]$ and new states as $\boldsymbol{o}_{t+1}=[o_1^{t+1},...,o_k^{t+1}]$ of all agents in the last bidding round. Store $(\boldsymbol{o}_t,\boldsymbol{a}_t,r_i^t,\boldsymbol{o}_{t+1})$ to replay buffer $D$. $\boldsymbol{o}_t\leftarrow \boldsymbol{o}_{t+1}$;
            \FOR{each agent $i = 1 \to k$}
                \STATE Sample $bs$ batches of samples $(\boldsymbol{o}_t,\boldsymbol{a}_i,r_i^t,\boldsymbol{o}_{t+1})$ from $D$;
                \STATE Update critic via optimizing loss defined in Eqs. (\ref{eq:9}-\ref{eq:10}), and update actor by Eq. (\ref{eq:11});
                \STATE Update target function parameters: $\theta_i^{Q^{t+1}}=\tau \theta_i^{Q^t}+(1-\tau) \theta_i^{Q^t}$,
                $\theta_i^{\boldsymbol{\varphi}^{t+1}}=\tau \theta_i^{\boldsymbol{\varphi}^t}+(1-\tau) \theta_i^{\boldsymbol{\varphi}^t}$;
            \STATE Store revenue of platform $\sum_{i=1}^k \sigma(S_i^t)$;
            \ENDFOR
        \ENDFOR
    \ENDFOR
    \RETURN{the maximum revenue of platform $\sum_{i=1}^k \sigma(S_i)$.}
\end{algorithmic}
\end{algorithm}

% \vspace{-0.5cm}
\subsection{Implementation}
It is difficult to save each state transition as a transition tuple in MBE like a typical Reinforcement Learning (RL) owing to the following reasons: (i) The states vary so frequently that it will take needless computation to save every bidding result of seed as a transition tuple.
(ii) The adopted CTDE paradigm requires merging all agents' state transitions and provides the merged transitions to agents for training. Therefore, we propose a Multi-agent Competitive Bidding Influence Maximization (MCBIM) algorithm with time complexity $O(NT(kl+kt_{up}+lt_{up}))$, where $t_{up}$ denotes the upper limit of diffusion step. 
At the beginning of the $t$-th bidding round, the actor $\varphi _i^t$ of each agent $i$ generates $l$ actions $\boldsymbol{a}_i^t$, then the agent $i$ executes $a_{ij}^t$ to bid for seed $s_j$ and get seed set $S_i$. After one bidding round, each agent $i$ will get a reward $r_i^t$ and update its state $o_i^t$ to new state $o_i^{t+1}$. Meanwhile, the platform will adjust the starting prices of all seeds by the bidding results in the last bidding round. Then we merge states and new states of all agents and store transition tuples and update network parameters with critic and actor update rules shown in Eqs. (\ref{eq:9}), (\ref{eq:10}), (\ref{eq:11}). More details are given in Algorithm \ref{algorithm:1}. 

Note that the state updates after all agents' actions have been executed, which means the frequencies of critic updates and actor executions are different. The critic updates occur once per bidding round, while each round involves $k$ agents independently generating $l$ actions to compete for the seeds. Therefore, for large-scale social networks in real-world scenarios, we can leverage existing industrial parallelization techniques (e.g., MapReduce) to simulate the action executions of $k$ agents synchronously, which can greatly improve the computing efficiency of MCBIM and enables MCBIM to solve the real-world problem with a limited time constraint. Taking MapReduce as an example, in the Map phase, we divide the $k$ agents into $k$ sub-tasks and assign them to different computing nodes. Each agent independently generates actions based on their policy and state, and executes them in parallel. In the Reduce phase, the execution results are merged to update the joint policies. The updated joint policies are then assigned to each agent for parallel updating of their individual policies in the next Map phase. Each bidding round can be accelerated through the above process. 
% This operation can improve the computing efficiency of MCBIM.

% \vspace{-0.5em}
\section{Experiments}\label{section:5}
To evaluate the effectiveness and efficiency of MCBIM, we compare it with four RL algorithms under the CLT model based on five datasets collected from real world. {\color{black}These algorithms are implemented using Python and the experiments are conducted on a machine with Ubuntu 16, an Intel Xeon Gold 6240R with 2.40GHZ CPU, 128GB memory, and an NVIDIA RTX 2080Ti GPU.} 

% \vspace{-0.5cm}
\subsection{Experimental Settings}
\textbf{Datasets.} The experiments are conducted on five real-world datasets. All datasets come from the Stanford  Dataset website \cite{snapnets}, and their detailed statistical information is shown in TABLE \ref{table:2}. Facebook is a network composed of friends lists of Facebook.com website. P2P-Gnutella09 is a snapshot of the Gnutella P2P network taken on August $9$, $2002$. Cit-HepPh is a paper citation network in Arxiv High Energy Physics category. DBLP is extracted from the DBLP Computer Science Bibliography, from which we select a collection of papers and their authors in the data mining field and build a collaboration network by connecting pairs of coauthors. Youtube is a video-sharing social network.

\begin{table}[htbp]
\scriptsize
\setlength{\abovecaptionskip}{0cm}  
\setlength{\belowcaptionskip}{-0.2cm}
\renewcommand{\arraystretch}{0.9}
\caption{Statistics of real-world networks}
\begin{center}
\begin{tabular}{llll}
\toprule
\textbf{Name}& \textbf{Type}& \textbf{Nodes}& \textbf{Edges}\\
\midrule
Facebook&Undirected&4,093&88,234\\
P2P-Gnutella09&Directed&8,114&26,013\\
Cit-HepPh&Directed&34,546&421,578\\
DBLP & Undirected&136,260&532,484\\
Youtube & Undirected&1,134,890&2,987,624\\
\bottomrule
\end{tabular}
\label{table:2}
\end{center}
% \vspace{-0.7cm}
\end{table}

\textbf{Parameter Settings.} For CLT model, {\color{black}the influence probability of each edge in directed graphs is defined as $w_{uv}=\frac{1}{\vert N^{in}(v)\vert},e_{uv}\in E$, while the influence probability in undirected graphs is defined as $w_{uv}=\frac{1}{\vert N(v)\vert},e_{uv}\in E$, where $N(v)$ is the degree of node $v$.} The activation threshold of each node is initialized by a random value between $0$ and $1$ \cite{li2020online,ghayour2021mlpr}. For MCBIM, we adopt the Adam optimizer and set the learning rate to $0.01$ and $\tau=0.01$ to update the target critic and target actor. $\gamma$ is set to $0.95$ and the size of replay buffer is set to $10^6$. After adding every $40$ transition tuples to the replay buffer, the target function parameters are updated. {\color{black}The batch size before making an update is set as $bs=1,024$ \cite{lowe2017multi}}. The constant $\omega$ in Eq. (\ref{eq:1}) is set to $2$ \cite{jamal2019task}, and the parameter $\kappa$ in Eq. (\ref{eq:4}) is set to $0.3$. We define one bidding round as an episode in our experiment, and the total number of episodes is equivalent to $T_b=T*N$, and $N$ is set to $50$ here.

\textbf{Comparison Algorithms.} 
We select four RL algorithms, DDPG \cite{DBLP:journals/corr/LillicrapHPHETS15}, TD3 \cite{fujimoto2018addressing}, SAC \cite{haarnoja2018soft} and PPO \cite{schulman2017proximal}, as comparison algorithms.
% and the information of algorithms are listed in Table \ref{table:1}, where the target policy and the action policy of the agent under off-policy are different but identical under on-policy, the distributed training decentralized execution (DTDE) paradigm indicates the agents have no extra information to train. 
All these four RL algorithms adopt distributed training decentralized execution (DTDE) paradigm (i.e., have no extra information for agents to train). 

\begin{list}{\labelitemi}{\leftmargin=1em}
\setlength{\itemsep}{0em}
\setlength{\parskip}{0pt}
\setlength{\parsep}{0pt}
    \item \textbf{DDPG}: DDPG is an off-policy learning algorithm based on DPG. It solves the continuous control problem by adopting Actor-Critic framework. (The off-policy indicates that the target policy and the action policy of the agent are different.)
    \item \textbf{TD3}: TD3 is an off-policy learning algorithm based on DPG. It improves the overestimation problem in DDPG by utilizing double Q-learning. 
    \item \textbf{SAC}: SAC is an off-policy learning algorithm to maximize the entropy-containing target. Different from DDPG and TD3, SAC optimizes the policy based on SPG.
    \item \textbf{PPO}: PPO is an on-policy learning algorithm based on SPG. It improves the Policy Gradient (PG) algorithm in the policy update strength by proposing a new objective function. (The on-policy indicates that the target policy and the action policy of the agent are identical.)
    \item \textbf{MCBIM} (ours): MCBIM is an off-policy learning algorithm based on DPG and uses the CTDE paradigm. The critic function takes joint action as input.
\end{list}

In the experiments, the agents that adopted the RL algorithms
are self-interested. They do not cooperate with the platform and only focus on maximizing their personal rewards in bidding.

\textbf{Performance Metrics.} The main metric we use for evaluation is the revenue of platform. Since CBIM has proven to be monotone (lemma \ref{lemma:3.2}), the optimal solution $\mathbb{S}^*$ should lie in the situation that all seeds have been bid successfully. However, when $k,l$ are relatively small, even Random algorithm has a high probability of achieving the situation that all seeds are bid successfully during thousands of episodes. For verifying this statement, we compare MCBIM with the above four algorithms and Random algorithm within $10^4$ episodes on Facebook, and the $k,l$ are set to $2$ and $5$ respectively. TABLE \ref{table:3} reports the results. Index $REV_{max}$ indicates the maximum revenue of platform while index $REV_{avg}$ indicates the average revenue of platform in all \emph{successful episodes} (i.e., the episode that all seeds are bid successfully), and indices $RE_i$ and $CR_i$ denote the average rewards and the average cost ratios (i.e., the ratio of cost to budget) of agent $i$ respectively. Index $SR$ indicates the ratio of successful episodes to the total episodes.

\begin{table}[htbp]
\scriptsize
\setlength{\abovecaptionskip}{0cm} 
\setlength{\belowcaptionskip}{-0.2cm} 
\renewcommand{\arraystretch}{0.8}
\caption{Bidding results of different algorithms on Facebook}
\begin{center}
\setlength{\tabcolsep}{0.8mm}{
\begin{tabular}{llllllll}
\toprule
\textbf{Indices} & $\boldsymbol{REV_{max}}$ &$\boldsymbol{REV}_{avg}$ &$\boldsymbol{RE_1}$  & $\boldsymbol{RE_2}$   & $\boldsymbol{CR_1}$   &  $\boldsymbol{CR_2}$    & $\boldsymbol{SR}$ \\
\midrule
\textbf{Random} &3,468.58 & 3,454.10 & 1,791.36 &  1,657.10 &  22.22\% & 21.40\% &0.13\% \\
\textbf{DDPG}  &3,470.16 & 3,460.96 & 1,978.94 & 1,544.16 & 81.37\% & 55.47\% & 5.93\% \\
\textbf{TD3 }& 3,476.12& 3,455.94 & 1,510.01&1,945.31&32.67\%&58.90\%&7.79\%\\
\textbf{SAC} &3,473.73&3,457.20&1,728.07&1,729.34&63.23\%&63.03\%&6.76\%\\
\textbf{PPO} &3,470.61 &3,457.61&1,676.43&1,781.08&60.10\%&63.30\%&6.15\%\\
\textbf{Ours} & 3,471.65&3,458.32&1,684.02&1,773.99&59.53\%&60.87\%&10.92\% \\
\bottomrule
\end{tabular}%
}
\label{table:3}
\end{center}
% \vspace{-0.4cm}
\end{table}

As TABLE \ref{table:3} shows, $REV_{max}$ and $REV_{avg}$ of different algorithms are almost identical, and the gap is the error in calculating the influence spread. Thus, it is difficult to make a comparison in the effectiveness of algorithms according to the revenue of platform. However, we notice that the $SR$ of algorithms is quite different which can reflect the probability of achieving the goal of CBIM, i.e., achieving a win-win situation.
% not only between the platform and the competitors, but also among competitors. 
Therefore, we utilize the $SR$ instead of the revenue of platform as the main metric to measure the effectiveness of algorithms when the problem scale (i.e., $k,l$) is small. 
Whether to use $SR$ or revenue of platform as the main metric can be empirically determined by observing values of both $SR$ and revenue of platform. When algorithms' $SR$s decrease to 0 and the revenue of platform obviously differs across algorithms along with the scale increases, we take revenue of platform for evaluation.
In addition, we utilize the running time to evaluate the efficiency of algorithms.

% \vspace{-0.5cm}
\subsection{ Results and Analyses}
\begin{figure*}[htbp]
    \centering     
    \includegraphics[width=\linewidth]{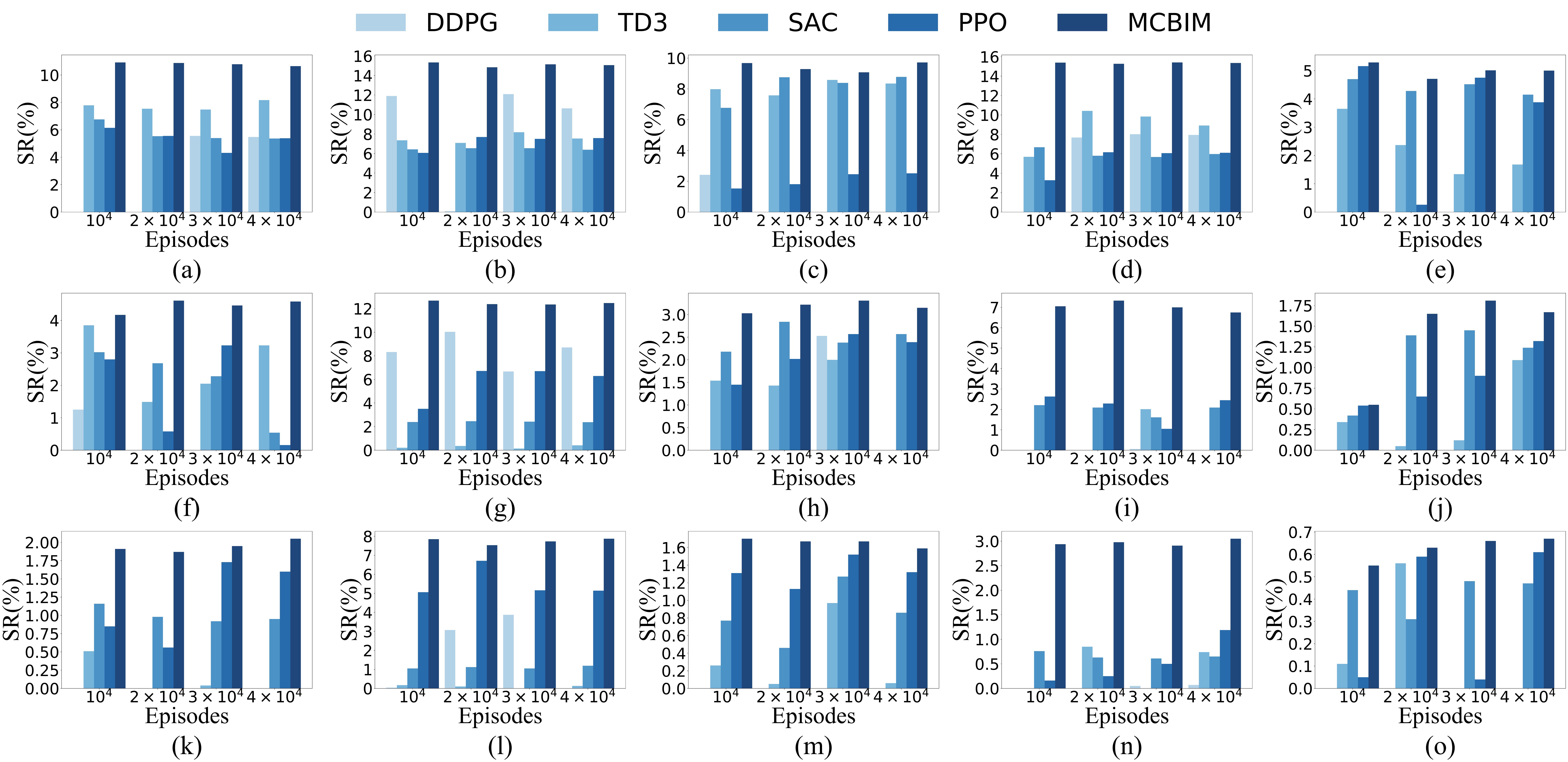}
    \caption{$SR$ for different number of episodes under different number of seeds.}
    \label{figure:4}
% \vspace{-0.4cm}
\end{figure*}

\textbf{Comparison under Different Number of Episodes and Seeds.} In the first set of experiments, we investigate the impact of varied amounts of episodes and seeds $l$ by changing episodes from $10^4$ to $4\times 10^4$ and $l$ from $5$ to $9$, while fixing the number of agents as $k=2$, the fairness threshold as $\rho=0.1$, and setting the same budgets $[\frac{l+1}{2},\frac{l+1}{2}]$ for both $2$ agents.

Fig. \ref{figure:4} shows the comparison results of $SR$ on five datasets under different episodes and different $l$ respectively. We can see that the $SR$ of MCBIM surpasses the other four RL algorithms in all cases, and MCBIM always has a much more stable performance on different datasets compared with the others. This observation between MCBIM and other algorithms demonstrates that taking all agents' actions and states as input to train the critic function for each agent is better than only taking its own actions and states as input for training. It also shows that MCBIM's way of cooperating with the platform yields a higher probability of reward guarantee for the competitors. 

In addition, comparing $SR$ on the same dataset with different $l$ as shown in Fig. 4(a), Fig. 4(f), and Fig. 4(k), we notice that $SR$ decreases as the growth of $l$. The reason lies in that the possible states of an agent grows exponentially along with $l$'s growth, and it is hard for the agent to learn the optimal policy $\pi$ through the state. However, this phenomenon of hard learning is more obvious in DDPG when $l=9$. The reason lies in that DDPG focuses on calculating the maximum value of $Q$ function, and it might lead to the overestimation of $Q$ value due to uncertain accurate maximum value.
Furthermore, we observe that the $SR$ of five algorithms on datasets with different topologies varies significantly as shown in Fig. \ref{figure:4}. The $SR$ that MCBIM can achieve on P2P-Gnutella09 is at least twice than that of other datasets. The reason is that the rewards returned by Youtube, Facebook, DBLP, and Cit-HepPh are dozens or even hundreds of times greater than that on P2P-Gnutella09, and the unit cost discrepancy of competitors on Youtube, Facebook, DBLP, and Cit-HepPh is much more significant than that on P2P-Gnutella09, which means the fairness constraint is more difficult to satisfy.

\begin{table}[htbp]
\scriptsize
\setlength{\abovecaptionskip}{0cm}  
\setlength{\belowcaptionskip}{-0.2cm}
\renewcommand{\arraystretch}{0.8}
\begin{center}
\caption{Running time for different number of episodes and seeds on Cit-HepPh (in seconds)}\label{table:4}
\setlength{\tabcolsep}{1mm}{
\begin{tabular}{lllllllllllll}
\toprule
$\boldsymbol{l}$  & \multicolumn{4}{c}{$\boldsymbol{5}$}         & \multicolumn{4}{c}{$\boldsymbol{7}$}         & \multicolumn{4}{c}{$\boldsymbol{9}$}         \\
\cmidrule{2-5}\cmidrule{6-9}\cmidrule{10-13}
\textbf{Episodes($\boldsymbol{10^4}$)} & $\boldsymbol{1}$ & $\boldsymbol{2}$ & $\boldsymbol{3}$ & $\boldsymbol{4}$ & $\boldsymbol{1}$ & $\boldsymbol{2}$ & $\boldsymbol{3}$ & $\boldsymbol{4}$ & $\boldsymbol{1}$ & $\boldsymbol{2}$ & $\boldsymbol{3}$ & $\boldsymbol{4}$ \\
\midrule
\textbf{DDPG}     & 30    & 98    & 190   & 323  & 46   & 134   & 269   & 447  & 56   & 172   & 343   & 555   \\
\textbf{TD3}      & 32    & 104    & 209   & 336   & 47    & 146   & 293   & 495   & 58    & 183   & -   & 592   \\
\textbf{SAC}      & 37    & 118    & 250   & 402   & 52    & 163   & 350   & 576   & 66    & 210   & 467   & 753  \\
\textbf{PPO}     & 44    & 122    & 235   & 378   & 55    & 159   & 312   & 516   & 69    & 200   & 401   & 665   \\
\textbf{Ours}   & {\color[HTML]{000000} \textbf{28}} & {\color[HTML]{000000} \textbf{79}} & {\color[HTML]{000000} \textbf{165}} & {\color[HTML]{000000} \textbf{264}} & {\color[HTML]{000000} \textbf{40}} & {\color[HTML]{000000} \textbf{113}} & {\color[HTML]{000000} \textbf{229}} & {\color[HTML]{000000} \textbf{377}} & {\color[HTML]{000000} \textbf{50}} & {\color[HTML]{000000} \textbf{144}} & {\color[HTML]{000000} \textbf{283}} & {\color[HTML]{000000} \textbf{466}}   \\
\bottomrule
\end{tabular}
}
\end{center}
% \vspace{-0.4cm}
\end{table}

Since the five algorithms' running time on five datasets are similar, we only show the results on Cit-HepPh. TABLE \ref{table:4} reports the results of running time under different $l$ and episodes on Cit-HepPh, where the empty value means the algorithm has not trained an effective policy satisfying fairness constraint and cannot work in testing. The five algorithms' running time increases along with the growth of both $l$ and episodes. MCBIM always runs faster than other algorithms. 
Specifically, compared with the second-fastest algorithm DDPG, MCBIM runs around $7\%\sim19\%$ faster than DDPG.
It demonstrates that using operations with various frequencies of critic updates and actor executions can improve the computing efficiency. In addition, we notice that the SAC's running speed decreases with the growth of episodes. The reason is that SAC conducts sampling with equal probability. Thus, the running speed of SAC decreases when the number of samples increases with the episodes.
The results in TABLE \ref{table:4} show that MCBIM strikes a good balance between efficiency and effectiveness.

\textbf{Comparison under Different Budgets.} In the second set of experiments, we investigate the impact of varied budgets. We fix $k=2$, $\rho=0.1$, and set $episodes=3\times 10^4$ based on the results of above experiments, 
%by comprehensively considering the performances of all algorithms under different episodes in the first group of experiments, 
then we set $2$ agents' budgets to $[3,2]$, $[4,3]$, $[5,4]$ corresponding to $l=[5,7,9]$ respectively. 

\begin{table}[t]
\scriptsize
\setlength{\abovecaptionskip}{0cm}  
\setlength{\belowcaptionskip}{-0.2cm}
\renewcommand{\arraystretch}{0.8}
\begin{center}
\setlength{\tabcolsep}{1pt}
\caption{$SR$ and running time for different competitor budgets on Facebook, P2P-Gnutella09 (abbreviated as P2P), Cit-HepPh (abbreviated as Cit), DBLP and Youtube ($RT$ in seconds)}\label{table:5}
\begin{tabular}{lcllllllllll}
\toprule
\multirow{2}{*}{\textbf{Dataset}}        & \multirow{2}{*}{\textbf{Budget}} & \multicolumn{5}{c}{$\boldsymbol{SR(\%)}$}                   & \multicolumn{5}{c}{$\boldsymbol{RT}$}              \\
\cmidrule{3-7}\cmidrule{8-12}
& & \textbf{DDPG}  & \textbf{TD3}   & \textbf{SAC}   & \textbf{PPO}   & \textbf{Ours} & \textbf{DDPG} & \textbf{TD3} & \textbf{SAC} & \textbf{PPO} & \textbf{Ours}\\
\midrule
\multirow{3}{*}{\textbf{Facebook}}& $\boldsymbol{[3,2]}$ & -  & 4.04  & 4.34  & 5.18  & \textbf{8.93}  & -  & 204 & 239 & 243 & \textbf{197} \\
& $\boldsymbol{[4,3]}$ & -  & 0.10  & 2.02  & 0.56  & \textbf{4.17}  & -    & 394 & 449 & 421 & \textbf{347} \\
& $\boldsymbol{[5,4]}$ & -  & 0.06  & 0.54  & 1.94  & \textbf{1.96}  & -    & 544 & 666 & 575 & \textbf{495} \\
\midrule
\multirow{3}{*}{\textbf{P2P}} & $\boldsymbol{[3,2]}$               & 15.05 & 21.24 & 25.96 & 19.39 & \textbf{26.25} & 191  & 200 & 229 & 234 & \textbf{178} \\
& $\boldsymbol{[4,3]}$ & 1.60  & 13.25 & 13.99 & 7.67  & \textbf{15.69} & 254  & 271 & 322 & 310 & \textbf{237} \\
& $\boldsymbol{[5,4]}$& 2.74  & -  & 9.06  & 3.17  & \textbf{10.70} & 315  & -   & 422 & 393 & \textbf{304} \\
\midrule
\multirow{3}{*}{\textbf{Cit}}  & $\boldsymbol{[3,2]}$  & 1.08  & 7.84  & 8.07  & 3.54  & \textbf{9.44}  & 167  & 191 & 221 & 220 & \textbf{164} \\
&  $\boldsymbol{[4,3]}$              & 0.60  & 2.78  & 2.82  & 1.89  & \textbf{3.47}  & 236  & 262 & 306 & 299 & \textbf{231} \\
& $\boldsymbol{[5,4]}$   & 0.98     & 0.04  & 0.37  & 0.37  & \textbf{1.75}  & 331  & 344 & 411 & 375 & \textbf{305} \\
\midrule
\multirow{3}{*}{\textbf{DBLP}}  & $\boldsymbol{[3,2]}$                & -  & 4.52  & 4.79  & 3.63  & \textbf{9.11}  & -    & 191 & 229 & 236 & \textbf{182} \\
& $\boldsymbol{[4,3]}$                  & -     & 1.46  & 1.49  & 1.19  & \textbf{3.61}  & -  & 276 & 326 & 320 & \textbf{256} \\
 & $\boldsymbol{[5,4]}$  & 1.14  & -  & 0.50  & 1.27  & \textbf{1.39}  & 341  & -   & 433 & 407 & \textbf{328}\\
\midrule
\multirow{3}{*}{\textbf{Youtube}}
& $\boldsymbol{[3,2]}$ & -  & 3.94  & 3.82  & 3.86  & \textbf{4.05}  & -    & 1,034  & 1,170 & 1,086 & \textbf{901} \\
& $\boldsymbol{[4,3]}$  & -     & 0.02 & 1.07  & 1.29  & \textbf{1.52}  & -  & 1,448 & 1,697 & 1,510 & \textbf{1,243} \\
 & $\boldsymbol{[5,4]}$  & -  & -  & 0.45  & 0.30  & \textbf{0.67}  & -  & -   & 2,291 & 2,008 & \textbf{1,678}\\
\bottomrule
\end{tabular}
\end{center}
% \vspace{-0.4cm}
\end{table}

TABLE \ref{table:5} reports the comparison results of $SR$ and $RT$ on five datasets with different competitor budgets, where index $RT$ indicates the running time (in seconds). MCBIM still surpasses the other algorithms according to the $SR$ and $RT$ on five datasets. Meanwhile, comparing results in TABLE \ref{table:5} with results in Fig. \ref{figure:4} under the same settings (i.e., same episodes, $l$ and dataset), we observe that the differences in the competitor budgets significantly affect $SR$. However, the variation tendency of $SR$ is different between P2P-Gnutella09, Cit-HepPh and Facebook, DBLP, Youtube, where $SR$ increases on P2P-Gnutella09, Cit-HepPh but decreases on Facebook, DBLP, Youtube. The reason also lies in the different order-of-magnitude rewards returned on datasets. The reward distributions on Facebook, DBLP and Youtube limit the fitting ability of agents' $Q$ functions, which leads to the biased evaluation of $Q$ values.

\textbf{Comparison under Different Number of Competitors and Seeds.} In the third set of experiments, we investigate the impact of varied amounts of competitors $k$ and seeds $l$ by changing $k$ from $5$ to $20$ and changing $l$ from $20$ to $100$, while fixing $episodes=4\times 10^4$, $\rho=0.3$, and then set the budget of each competitors as $\frac{l}{2}$. With the expansion of the problem scale, the $SR$ decreases to 0, we thus utilize revenue of platform instead of $SR$ as the main metric to measure the performance of algorithms. 

\begin{table}[t]
\scriptsize
\setlength{\abovecaptionskip}{0cm}  
\setlength{\belowcaptionskip}{-0.2cm}
\renewcommand{\arraystretch}{0.8}
\begin{center}
\caption{Results under different number of competitors and seeds on P2P-Gnutella09 ($RT$ in seconds)}\label{table:6}
\begin{tabular}{llllll}
\toprule
\multirow{2}{*}{$\boldsymbol{k,l}$}   & \multirow{2}{*}{\textbf{Indices}} & \multicolumn{4}{c}{\textbf{Algorithms}} \\
\cmidrule{3-6}
& & \textbf{DDPG} & \textbf{SAC} & \textbf{PPO} & \textbf{Ours}\\
\midrule
\multirow{3}{*}{$\boldsymbol{5,20}$} 
& $\boldsymbol{SER(\%)}$ &35.09 & 90.47 & 21.40 & \textbf{91.97}\\
&$\boldsymbol{RT}$ &3,273 & 3,996 & 5,103 & \textbf{1,295} \\
& $\boldsymbol{ROP_{max}}$& 605.073 & 650.255 & 656.471 &
\textbf{697.196} \\
\midrule
\multirow{3}{*}{$\boldsymbol{10,50}$} 
& $\boldsymbol{SER(\%)}$ &52.87 & 92.09 &1.15 & \textbf{93.64}\\
&$\boldsymbol{RT}$ &13,750 & 16,136 & 22,880 & \textbf{5,624} \\
& $\boldsymbol{ROP_{max}}$& 841.882 & 900.016 & 913.776 &
\textbf{953.231} \\
\midrule
\multirow{3}{*}{$\boldsymbol{20,100}$} 
& $\boldsymbol{SER(\%)}$ &81.80 &97.62 & 1.59 & \textbf{99.04}\\
&$\boldsymbol{RT}$ &27,228 & 31,039 & 47,680 & \textbf{11,294} \\   
& $\boldsymbol{ROP_{max}}$& 991.948 & 1,077.968 & \textbf{1,118.330} &
1,091.947 \\
\bottomrule
\end{tabular}    
\end{center}
\vspace{-0.4cm}
\end{table}

Owing to the length constraint, we only show the comparison results on P2P-Gnutella09 and the results of MCBIM on five datasets. TABLE \ref{table:6} reports the comparison results of $SER$, $RT$, and $ROP_{max}$ on P2P-Gnutella09 under different $k$ and $l$, where index $SER$ refers to the ratio of the episodes satisfying the fairness constraint to total episodes, and it can reflect the ability of algorithm in fairness guarantee, index $ROP_{max}$ indicates the maximum revenue of platform in the episodes while satisfying the fairness constraint. 
Since the results of TD3 cannot satisfy the fairness constraint and has no results on five datasets, we thus do not display TD3 in TABLE \ref{table:6}. From the obtained results, we can see that the $SER$, $RT$, $ROP_{max}$ of MCBIM outperform the other three algorithms in most cases, where the high $SER$ of MCBIM shows its excellent capability in keeping fairness among competitors. Meanwhile, the shortest $RT$ of MCBIM again demonstrates the good efficiency of the operation of critic updates and actor executions with different frequencies. In addition, although the $ROP_{max}$ of PPO under $k=20,l=100$ is higher than that of MCBIM, PPO's running time is always the longest among four algorithms. Moreover, the extreme low $SER$ indicates that it is hard for PPO to safeguard the fairness.

\begin{table}[htb]
\scriptsize
\setlength{\abovecaptionskip}{0cm}  
\setlength{\belowcaptionskip}{-0.2cm}
\renewcommand{\arraystretch}{0.8}
\begin{center}
\caption{Results of MCBIM under different number of competitors and seeds ($RT$ in seconds)}\label{table:7}
\begin{tabular}{lllll}
\toprule
\multirow{2}{*}{\textbf{Dataset}}   & \multirow{2}{*}{$\boldsymbol{k,l}$} & \multicolumn{3}{c}{\textbf{Indices}} \\
\cmidrule{3-5}
                                    &                        & $\boldsymbol{SER(\%)}$     & $\boldsymbol{RT}$ & $\boldsymbol{ROP_{max}}$   \\
\midrule
\multirow{3}{*}{\textbf{Cit-HepPh}} & $\boldsymbol{5,20}$                 & 90.91    & 1,207   & 6,289.162    \\
& $\boldsymbol{10,50}$& 81.95  & 5,149  & 7,552.997   \\
 &$\boldsymbol{20,100}$ & 96.79     & 10,398      &8,579.818    \\
\midrule
\multirow{3}{*}{\textbf{Facebook}}  & $\boldsymbol{5,20}$ & 42.16       & 1,562   & 2,024.389      \\
&$\boldsymbol{10,50}$& 60.61   & 6,448   & 2,756.884    \\
& $\boldsymbol{20,100}$ & 52.19  & 13,896   & 3,270.420  \\
\midrule
\multirow{3}{*}{\textbf{DBLP}}     & $\boldsymbol{5,20}$& 94.49  & 1,382    & 3,925.268    \\
& $\boldsymbol{10,50}$  & 99.87   & 6,200  & 6,439.803     \\
&$\boldsymbol{20,100}$   & 99.97  & 12,725  & 9,141.457       \\  
\midrule
\multirow{3}{*}{\textbf{Youtube}}
& $\boldsymbol{5,20}$ & 38.06  &  2,815  & 106,842.017   \\
& $\boldsymbol{10,50}$  & 35.52  & 11,193 & 111,890.591  \\
& $\boldsymbol{20,100}$  & 30.60 & 22,436 & 119,381.433 \\  
\bottomrule
\end{tabular}    
\end{center}
% \vspace{-0.4cm}
\end{table}

TABLE \ref{table:7} reports the results of MCBIM under different competitors and seeds on the remaining four datasets. MCBIM still keeps high probability in keeping fairness on datasets except Facebook and Youtube. 
%We notice that the $SER$ of MCBIM on dataset Facebook is much lower compared to the other datasets. 
The relatively low $SER$ on Facebook and Youtube is because the discrepancies of the influence spreads among seeds on Facebook and Youtube are significantly large, which leads to the big differences in the unit cost distribution among competitors, and it is more difficult to meet the fairness constraint compared to the other three datasets. 
In addition, we observe that the difference of $RT$ among five datasets except Youtube is small. The reason lies in that the reward computation we utilize is a degree-centrality algorithm which can return the reward to the environment quickly. Therefore, the time for calculating a reward is similar for datasets other than Youtube which is still time-consuming. Moreover, the $RT$ increases as the problem scale grows. The reason is that the state dimension of competitors increases proportionally with the growth of $k$ and $l$, so that the running time spent on action computing increases with the state dimension. In summary, the experimental results demonstrate that MCBIM has good scalability.

\begin{table}[t]
\scriptsize
\setlength{\abovecaptionskip}{0cm} 
\setlength{\belowcaptionskip}{-0.2cm} 
\renewcommand{\arraystretch}{0.8}
\caption{Results under different ratios of $k:l$ on P2P-Gnutella09 ($RT$ in seconds)}
\begin{center}
\begin{tabular}{llll}
\toprule
$\boldsymbol{k:l}$ &$\boldsymbol{SER}$&$\boldsymbol{RT}$  & $\boldsymbol{ROP_{max}}$ \\
\midrule
$\boldsymbol{20:100}$ &99.04 & 11,294 & 1,091.947 \\
$\boldsymbol{40:100}$  &15.00 & 14,346 & 730.873 \\
$\boldsymbol{60:100}$& -& - & -\\
\bottomrule
\end{tabular}
\label{table:8}
\end{center}
\vspace{-0.4cm}
\end{table}

\textbf{Scaling Study on the Number of Competitors and Seeds.} In the fourth set of experiments, we investigate the impact of ratio $k:l$ on our algorithm MCBIM, and the comparison results are shown as follows: (1) Keep $k$ unchanged but enlarge $l$. We can observe the impact of $k:l$ on the results shown in the first group of experiments, where we test ratios of $k:l=[2:5,2:7,2:9]$. As shown in Fig. \ref{figure:4}, we can observe that the $SR$ decreases as the ratio decreases. This is because a lower ratio of $k:l$ results in a larger gap between competitors’ seed sets, leading to a higher frequency of situations with serious disparities across competitors and a lower probability of achieving fairness. (2) Keep the $l$ unchanged but enlarge $k$. We conduct experiments on ratios $k:l=[20:100,40:100,60:100]$ to explore the impact of $k:l$, and we only show the results on dataset P2P-Gnutella09 due to the space limitation. The results are shown in TABLE \ref{table:8}, which can also reflect the performance of our proposed model when the $k$ value scales up. From the results shown in TABLE \ref{table:8}, we can observe that both the $SER$ and $ ROP_{max}$ decrease as the ratio (or $k$) increases. The decrease in $SER$ is due to the more intense competition resulting from the increase of $k$, making the fairness among competitors more difficult to achieve. The decrease in $ROP_{max}$ is because the competitive seed sets propagated in the networks increase with the increase of $k$, making the sum of the influence probability of each competitor’s activated nodes toward an inactive node decrease, which reduces the reward of each competitor. Therefore, the revenue of platform decreases accordingly. (3) Enlarge both $k$ and $l$. We can observe the impact of $k:l$ on results shown in the third group of experiments, where we test the ratios of $k:l=[5:20,10:50]$ (or $[5:20,20:100]$). As shown in TABLE \ref{table:6} and TABLE \ref{table:7}, we can observe that the impact of $k:l$ on the metric $SER$ is small. The reason is that the increase of $l$ effectively relieves the fierce competition atmosphere resulting from the growth of $k$.

% \vspace{-0.5em}
\section{Conclusions}\label{section:6}
We investigated a new IM problem, Competitive Bidding Influence Maximization (CBIM) in this work. The goal of CBIM is to achieve a win-win situation both between the platform and the competitors and among competitors on the premise of fairness through multi-round bidding. We showed that the CBIM problem is proven to be NP-hard. We put forward a novel Fairness-aware Multi-agent Competitive Bidding Influence Maximization (FMCBIM) framework to address the CBIM problem. In FMCBIM framework, we built a Multi-agent Bidding Particle Environment (MBE) to capture the interactions among competitors during the dynamic bidding process. Furthermore, we proposed a practical Multi-agent Competitive Bidding Influence Maximization (MCBIM) algorithm by leveraging the MARL technique to safeguard rewards and fairness for competitors. A large number of experiments were conducted on five real-world datasets, which proved the effectiveness of our proposed framework and the superiority of the MCBIM algorithm over four comparison algorithms.

% \vspace{-0.5em}
%\section*{Acknowledgment}
%This work was partially supported by the National Natural Science Foundation of China (No. 61972272). 

% \vspace{-0.5em}
\bibliographystyle{IEEEtran}
\bibliography{reference}

% \newpage
% \centerline{FOOTNOTES}

% \footnote{}{This work was partially supported by the National Natural Science Foundation of China under Grant 61972272, the Natural Science Foundation of the Jiangsu Higher Education Institutions of China under Grant 21KJA520008, and Qinlan Project of Jiangsu Province.}

% \clearpage

% \newpage
% {\renewcommand*\numberline[1]{Fig.\,#1:\space}
% \makeatletter
% \renewcommand*\l@figure[2]{\noindent#1\par}
% \makeatother

% \listoffigures}
% \vspace{-48em}
\begin{IEEEbiography}[{\includegraphics[width=1in,height=1.25in,clip,keepaspectratio]{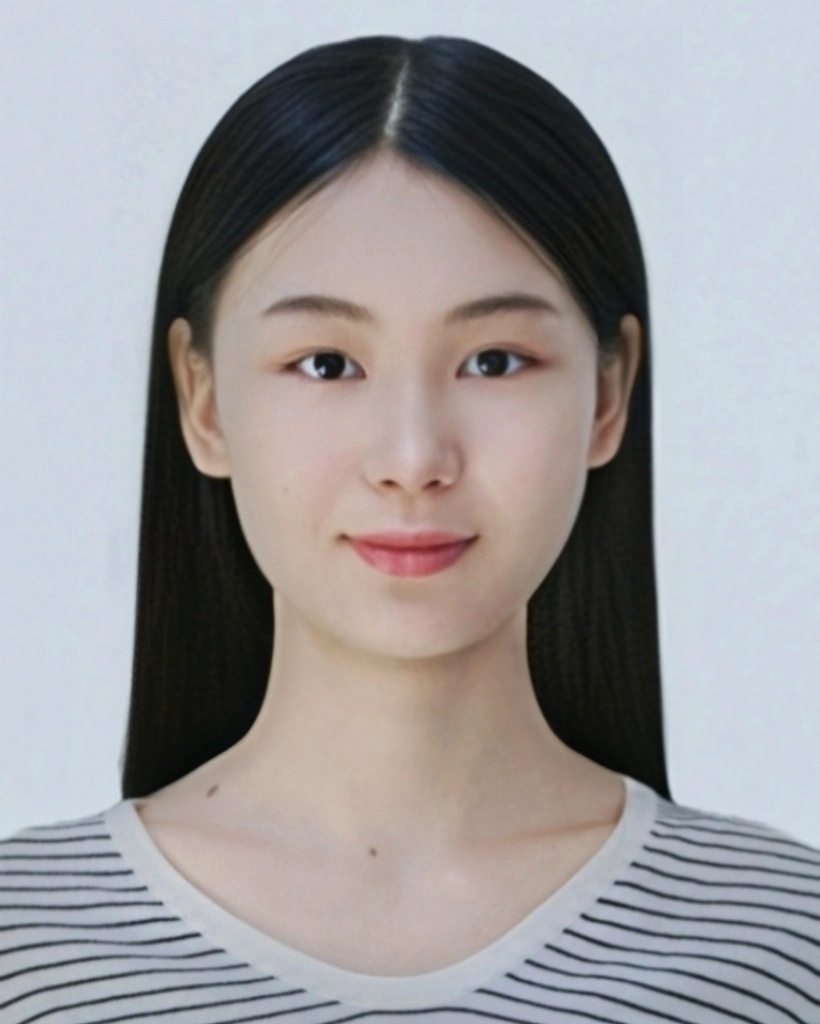}}]{Congcong Zhang}
received the B.S. degree in computer science from Anhui Normal University, Wuhu, China, in 2021. She is currently pursuing the M.S. degree with Soochow University, Suzhou, China. Her research interests include online social networks and social computing. 
\end{IEEEbiography}

\begin{IEEEbiography}
[{\includegraphics[width=1in,height=1.25in,clip,keepaspectratio]{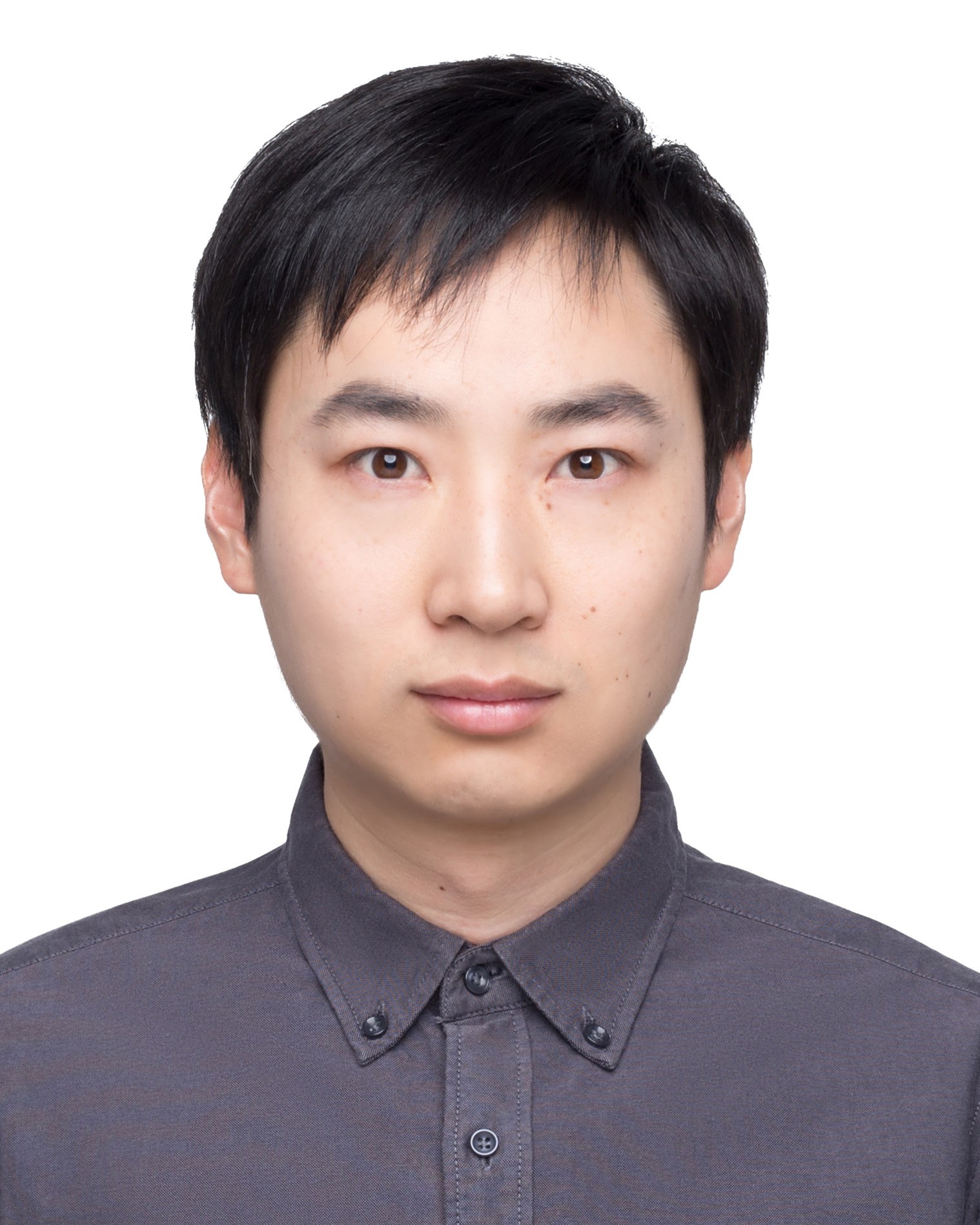}}]{Jingya Zhou} (Member, IEEE)
received the B.E. degree in computer science from Anhui Normal University, Wuhu, China, in 2005, and the Ph.D. degree in computer science from Southeast University, Nanjing, China, in 2013. He is currently an associate professor
with the School of Computer Science and Technology, Soochow University, Suzhou,
China. He has co-authored more than 70 papers in these areas, many of which have been published in well-known journals and conferences such as ACM CSUR, IP\&M, WWW, INFOCOM, ICDE, ICDCS, ICPP, and DASFAA. His research interests include cloud and edge computing, network embedding, online social networks, and data center networking.
\end{IEEEbiography}

\begin{IEEEbiography}
[{\includegraphics[width=1in,height=1.25in,clip,keepaspectratio]{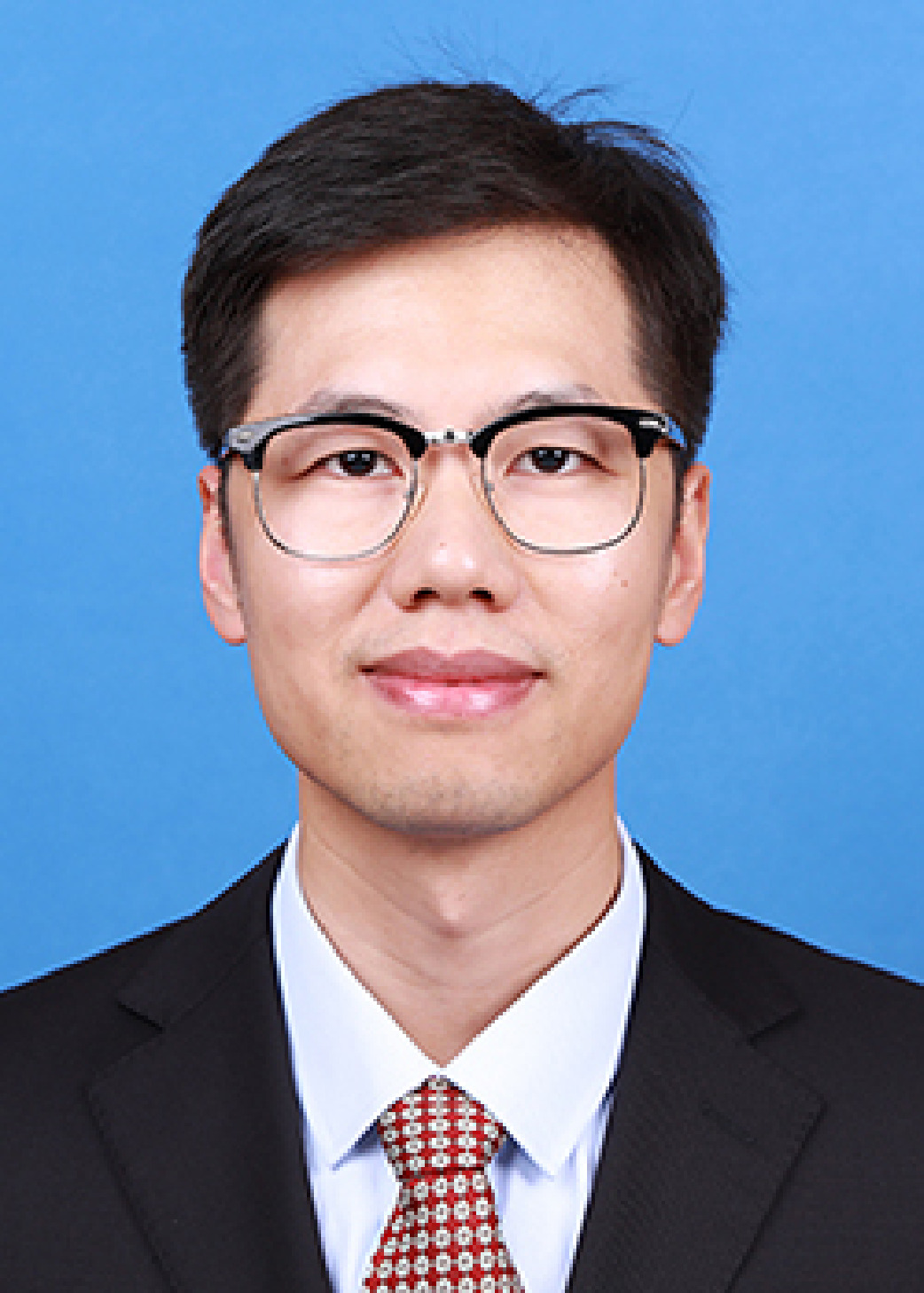}}]{Jin Wang} (Member, IEEE)
received the B.S. degree from Ocean University of China in 2006, and the Ph.D. degree in computer science jointly awarded by City University of Hong Kong and University of Science and Technology of China in 2011. He is currently a professor at the Department of Computer Science and Technology, Department of Future Science and Technology, Soochow University, Suzhou, China. His research
interests include edge computing and network security. 
\end{IEEEbiography}

\begin{IEEEbiography}
[{\includegraphics[width=1in,height=1.25in,clip,keepaspectratio]{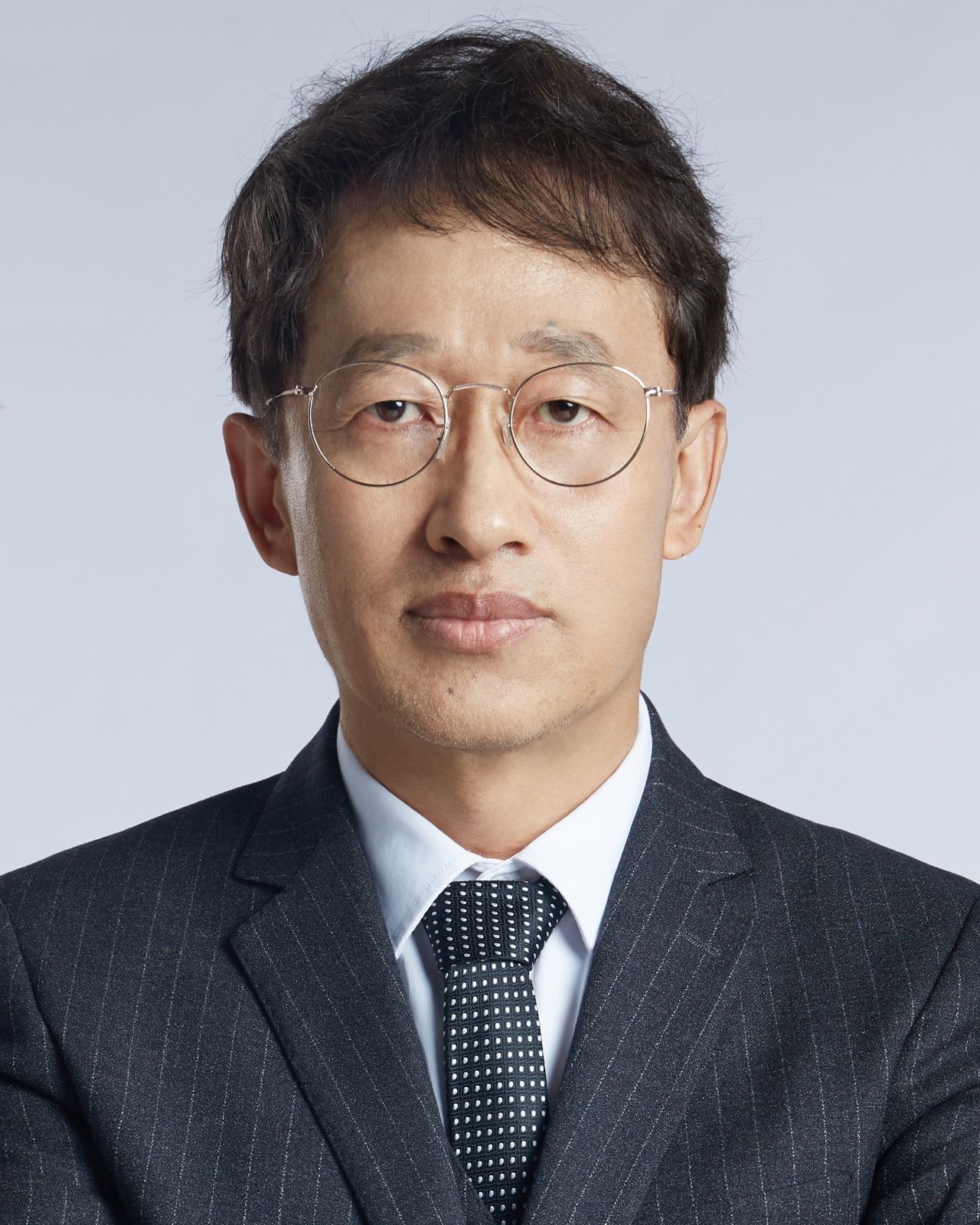}}]{Jianxi Fan}
received the B.S. degree in computer science from Shandong Normal University, Jinan, China, in 1988, the M.S. degree in computer science from Shandong University, Jinan, China, in 1991, and the Ph.D. degree in computer science from the City University of Hong Kong, Hong Kong, in 2006. He is currently a Professor with the School of Computer Science and Technology, Soochow University, Suzhou, China. He has authored or co-authored more than 150 research papers in refereed journals and conferences, including IEEE TPDS, IEEE TC, IEEE TDSC, IEEE TSMC, ACM Comp. Survey, INFOCOM, and ICDCS. His research interests include parallel and distributed systems, interconnection architectures, data center networks, and graph theory.
\end{IEEEbiography}

\begin{IEEEbiography}
[{\includegraphics[width=1in,height=1.25in,clip,keepaspectratio]{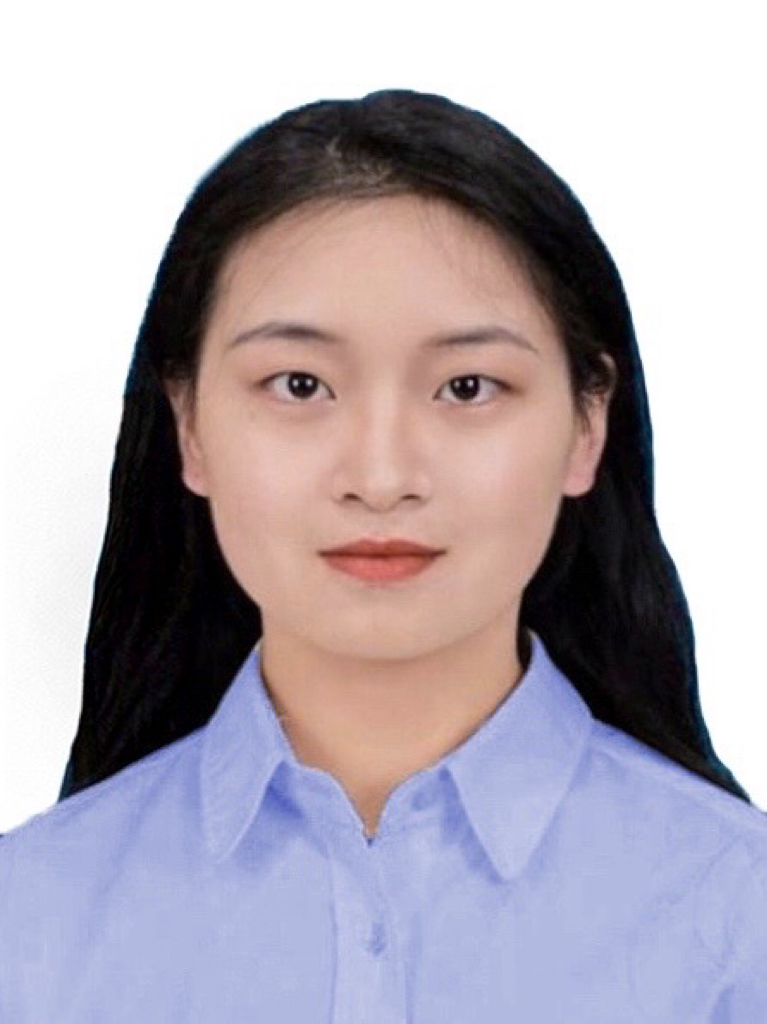}}]{Yingdan Shi}
received the B.S. degree in computer science from Hangzhou Normal University, Hangzhou, China, in 2021. She is currently pursuing the M.S. degree with Soochow University, Suzhou, China. Her research interests include online social networks and social computing.
\end{IEEEbiography}
\end{document}